\documentclass[useAMS, usenatbib]{mn2e}

\makeatother
\usepackage[normalem]{ulem} 

\usepackage{epsfig,amssymb,amsmath,bm} 
\usepackage{graphicx,epstopdf}
\usepackage{amsfonts, cancel}
\usepackage{marvosym}
\usepackage{dsfont}
\usepackage{stmaryrd}
\usepackage{color}
\usepackage{subfigure}
\usepackage{verbatim}
\usepackage{dcolumn}
\usepackage{latexsym}
\usepackage{subfigure}
\usepackage{mathtools}
\usepackage[english]{babel}

\newcommand{\be}{\begin{equation}}
\newcommand{\ee}{\end{equation}}
\newcommand{\f}{\frac}

\newcommand{\ball}{\begin{align}}
\newcommand{\eall}{\end{align}}
\newcommand{\bsy}{\boldsymbol}

\title[Simulating the frequency dependence of scattering imprints on pulsar observation]{The frequency dependence of scattering imprints on pulsar observations}
\title[Simulating the frequency dependence of scattering imprints on pulsar observation]{The frequency dependence of scattering imprints on pulsar observations}
\author[M Geyer and A Karastergiou]{M. Geyer$^{1,3}$\thanks{E-mail: marisa.geyer@physics.ox.ac.uk} and A.
Karastergiou$^{1,2,3}$\\
$^{1}$Astrophysics, University of Oxford, Denys Wilkinson Building, Keble Road, Oxford OX1 3RH, UK\\
$^{2}$Physics Department, University of the Western Cape, Cape Town 7535, South Africa\\ $^{3}$Department of Physics and Electronics, Rhodes University, PO Box 94, Grahamstown 6140, South Africa}
\begin{document}

\date{Accepted 14 July 2016. Received 11 July 2016; in original form 14 March 2016.}

\pagerange{\pageref{firstpage}--\pageref{lastpage}} \pubyear{2016}

\maketitle

\label{firstpage}

\begin{abstract}
Observations of pulsars across the radio spectrum are revealing a dependence of the characteristic scattering time ($\tau$) on frequency, which is more complex than the simple power law with a theoretically predicted power law index. In this paper we investigate these effects using simulated pulsar data at frequencies below 300 MHz. We investigate different scattering mechanisms, namely isotropic and anisotropic scattering, by thin screens along the line of sight, and the particular frequency dependent impact on pulsar profiles and scattering time scales of each.  We also consider how the screen shape, location and offset along the line of sight lead to specific observable effects. We evaluate how well forward fitting techniques perform in determining $\tau$. We investigate the systematic errors in $\tau$ associated with the use of an incorrect fitting method and with the determination of an off-pulse baseline. Our simulations provide examples of average pulse profiles at various frequencies. Using these we compute spectra of $\tau$ and mean flux for different scattering setups. We identify setups that lead to deviations from the simple theoretical picture. This work provides a framework for interpretation of upcoming low frequency data, both in terms of modelling the interstellar medium and understanding intrinsic emission properties of pulsars.

\end{abstract}


\section{Introduction}
Observed radio pulses from pulsars pass through the ionised
interstellar medium (ISM) before terrestrial detections are made. As
the waves propagate through the ISM, inhomogeneities in the electron
densities of dense, ionised regions cause the radio waves to scatter.
This scattering is observed through, amongst other effects, the
frequency dependent temporal broadening of the received radio pulses.
The temporal broadening typically leads to characteristic exponential \textit{tails}
in averaged pulsar profiles (e.g. \citealt{Lohmer2001}). Understanding the frequency dependence of the temporal broadening allows us to investigate the properties of the scattering media. Compensating for scattering effects can be used to infer the intrinsic radio emission characteristics of pulsars.

The scattering tails of the broadened profiles are routinely modelled with exponential functions that take the form $\tau^{-1}e^{-t/\tau}$, where $\tau$ is referred to as the \textit{characteristic scattering time}. This functional form, as described in detail in Sec. \ref{sec:scatsetup}, is based on the assumption of a single, thin scattering screen extending infinitely transverse to the line of sight to the pulsar. The screen is invoked as an approximation of the combined scattering that the pulse signal undergoes in the extended ISM along the line of sight \citep{Scheuer1968, Cronyn1970,Williamson1972, Williamson1973}. The characteristic scattering time $\tau$ depends on the properties of the scattering screen as well as its location along the line of sight (see eq. \ref{eq:tau}) and it is maximised for a midway screen. 

Apart from approximating the overall scattering geometry to a single scatterer, historic studies have also investigated distinct electron density distribution models for the screen. These include an isotropic Gaussian electron column density wavenumber spectrum, which leads to a temporal broadening dependence on frequency, $\tau \propto \nu^{-4}$, \citep{LeeJokipii1976, Lang1971}. Alternatively power law wavenumber spectra, characterised by an inner and outer cut-off length scale, are adopted to model the plasma inhomogeneities. Most often a Kolmogorov power spectrum, which describes the turbulence of a neutral gas, is preferred \citep{Rickett1977, RCB1984, Rickett1990}. A Kolmogorov spectrum will have a frequency dependence of $\tau \propto \nu^{-4.4}$, for wavenumbers lying between the inner and the outer cut-off length scales\footnote{Note that for power-law spectra, where $\tau \propto \nu^{-\alpha}$, the fluctuation spectral index $\beta$ is often quoted instead of the spectral index $\alpha$. In the 2D case, for $\alpha \geq 4$, they are related through: $\alpha = 2\beta/(\beta -2)$. A Kolmogorov spectrum with $\alpha = 4.4$ has $\beta = 11/3.$ Also note that the symbols $\alpha$ and $\beta$ are sometimes used interchangeably.}.

The scale at which electron density inhomogeneities lead to strong diffractive scintillation and consequently the observed pulse broadening analysed in this paper, is set by the field coherence scale, $s_{0} = 1/(k \theta)$, with $k$ the radio wavenumber \citep{Rickett1990}. For strong diffractive scintillation $s_{0} \ll L_{F}$, where $L_{F} = (D/k)^{1/2}$, is the Fresnel scale (e.g. \citealt{SmithThompson1988}. Refer to Fig. \ref{GeoGraph} for descriptions of $\theta$ and $D$.).

Estimations of $\tau$ values from pulsar profile observations are done by either deconvolution of the folded pulse profile with a broadening function such as a one-sided exponential function through e.g. the $CLEAN$-based algorithms \citep{Bhat2003CLEAN}, or by forward modelling which convolves a template unscattered pulse profile (or alternatively an observed high frequency pulse profile) with a parametrised broadening function and compares the model to the observed profile (e.g. \citealt{Lohmer2001,Lohmer2004}). It should be noted that using a high frequency observation as a template will, for pulsars that exhibit intrinsic frequency evolution, introduce biases to measurements of scattering parameters. A deconvolution method does not suffer from this effect.

Alternatively $\tau$ can be estimated by means of the \textit{scintillation bandwidth}, $\delta f$ \citep{LyneRickett1968}. The scattering process imparts frequency dependent phase changes on the passing radio waves. An interference pattern will emerge for phase changes of the order $2\pi \delta f \tau \approx 1$ radian. A measurement of $\delta f$, over which the interference effects are observed in the dynamic spectrum, can therefore lead to an approximation of $\tau$ \citep{Cordes1986}.

Using the ISM transfer function implied by the thin screen model, results in fits of $\tau$ in observed data that follow the expected spectral dependence \citep{Armstrong1995}. However, deviations from the theoretical spectral dependences have also been proposed and observed - both with spectral index $\alpha$ (where $\tau \propto \nu^{-\alpha}$) smaller than predicted \citep{Bhat2004, Lohmer2001, Lewandowski2015} and larger than predicted \citep{Rickett1990, LambertRickett1999,Tuntsov2012}.
Different models will lead to different values for $\alpha$, but the critical minimum for strong diffractive scattering, is $\alpha = 4$ \citep{RNB1986}.

Anisotropic scatterers, i.e. media for which the distribution of scattering angles exhibit directionality have been considered as causes for steeper spectra \citep{Tuntsov2012}. They were first invoked to explain organised patterns in the dynamic spectra of pulsar observations \citep{Gupta1994}. \citet{Stinebring2001} found that complex patterns in dynamic spectra are often related to parabolic arcs in the secondary (power) spectra of pulsar observations, which can be explained by anisotropic scattering \citep{Walker2004}. There has also been some successes in associating anisotropy with observations of Intra Day Variability (IDV) \citep{Bignalletal2003, Tuntsov2012}. Extreme Scattering Events (ESE's) as observed by \citet{Fiedler1987} and analysed by e.g. \citet{Walker2006} have motivated works studying the impact of alternative scatterers, such as self-gravitating, AU-sized ionised gas clouds. A one dimensional Kolmogorov power spectrum in electron density distribution is often used to model extreme anisotropy. 

On the contrary, weaker dependences on frequency have been attributed to screens which have a limited physical size rather than being infinite transverse to the line of sight as most models assume \citep{CordesLazio2001}. Beyond the edges of such a screen, no radio waves coming from the pulsar will be bent back into the line of sight of the observer, leading to no further scatter broadening with increasing wavelength. Such configurations also lead to a loss in observable flux.  

A physical inner scale of the scattering screen, that is larger than the diffractive scattering scale,
will restrict diffraction to a maximum scattering angle. This too will lead to less power observed at large angles, or alternatively flatter $\tau$ spectra with lower power law indices, at long wavelengths. In the case of an inner scale however, flux is preserved \citep{RickettJohnston2009}.

Until recently, characterising the frequency dependence of temporal scatter broadening was done using pulsar observations taken at different receivers and telescopes at different epochs. With the construction of telescopes such as the Low-Frequency Array (LOFAR; \citealt{vanHaarlem2013}) and the Murchison Widefield Array (MWA; \citealt{Tingay2012}), we have the advantage of studying pulsars at low frequencies, where scattering is most pronounced, using broad bands (see \citealt{Stappers2011} for an outline of LOFAR's pulsar observing modes). This allows accurate measurements of $\tau$ spectra and simultaneous studies of the flux spectrum and profile evolution of the pulsar. Analysis of broad band data will therefore test our current understanding of pulsar scattering. 

Recent works showing average pulse profiles in the LOFAR bands (30 - 90 MHz and 110 - 190 MHz) confirm the expectation of scatter broadened pulse shapes at these frequencies \citep{Bilous2015, Pilia2015}.  

In this paper we provide a framework based on simulations, for interpreting observations such as the ones presented in the LOFAR works. This can be used to gain insight on the size, location and nature
of scattering screens.

The paper is split into sections as follows. In Sec.~\ref{sec:simdataandtech} we describe the general scattering setup, the simulated data and the techniques by which we fit the data to retrieve $\tau$ spectra. The mathematical formalism for examples of the broadening functions for isotropic and anisotropic scattering is also presented in this section. In Sec.~\ref{sec:results} we present the results of our experiments. These include example profile shapes as well as $\tau$ and flux spectra for infinite and finite screens respectively. We end with a discussion of our results and their relevance to broad band observations in Sec.~\ref{sec:discussion}, followed by a short conclusion. 

\section{Simulated Data and Fitting Techniques} \label{sec:simdataandtech}
\subsection{Scattering geometry}\label{sec:scatsetup}
\begin{figure}
\centering
\includegraphics[width = \columnwidth]{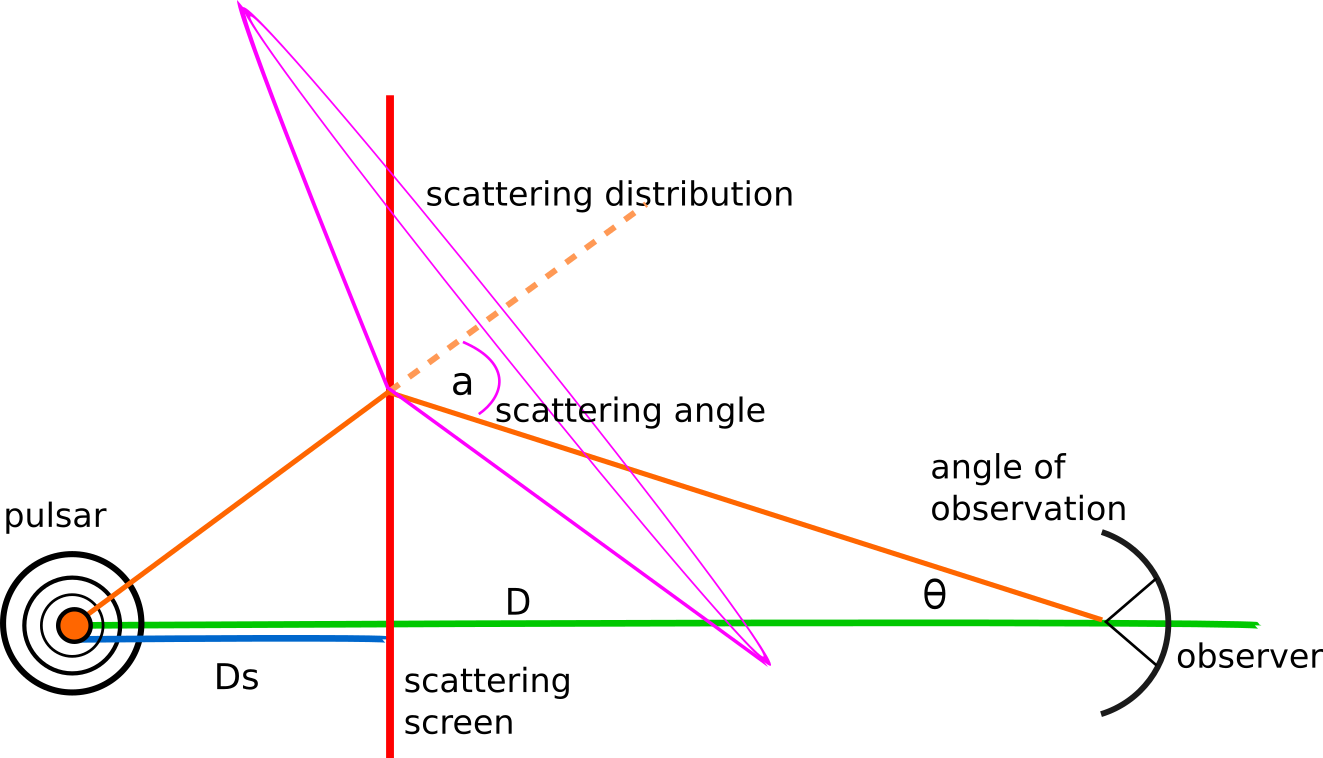}
\caption{A schematic representation of radio waves scattered by a single scattering screen, with ${D_s}$ the distance from the pulsar to the screen and $D$ the overall distance from the pulsar to the observer. The scattering angle $a$, (drawn from some probability distribution, here represented by the magenta cone), is the angle by which the ray is scattered away from its straight line trajectory and $\theta$ the angle at which we observe the scattered ray.}
\label{GeoGraph}
\end{figure}

\begin{figure*}
\centering
\includegraphics[width=0.8\textwidth]{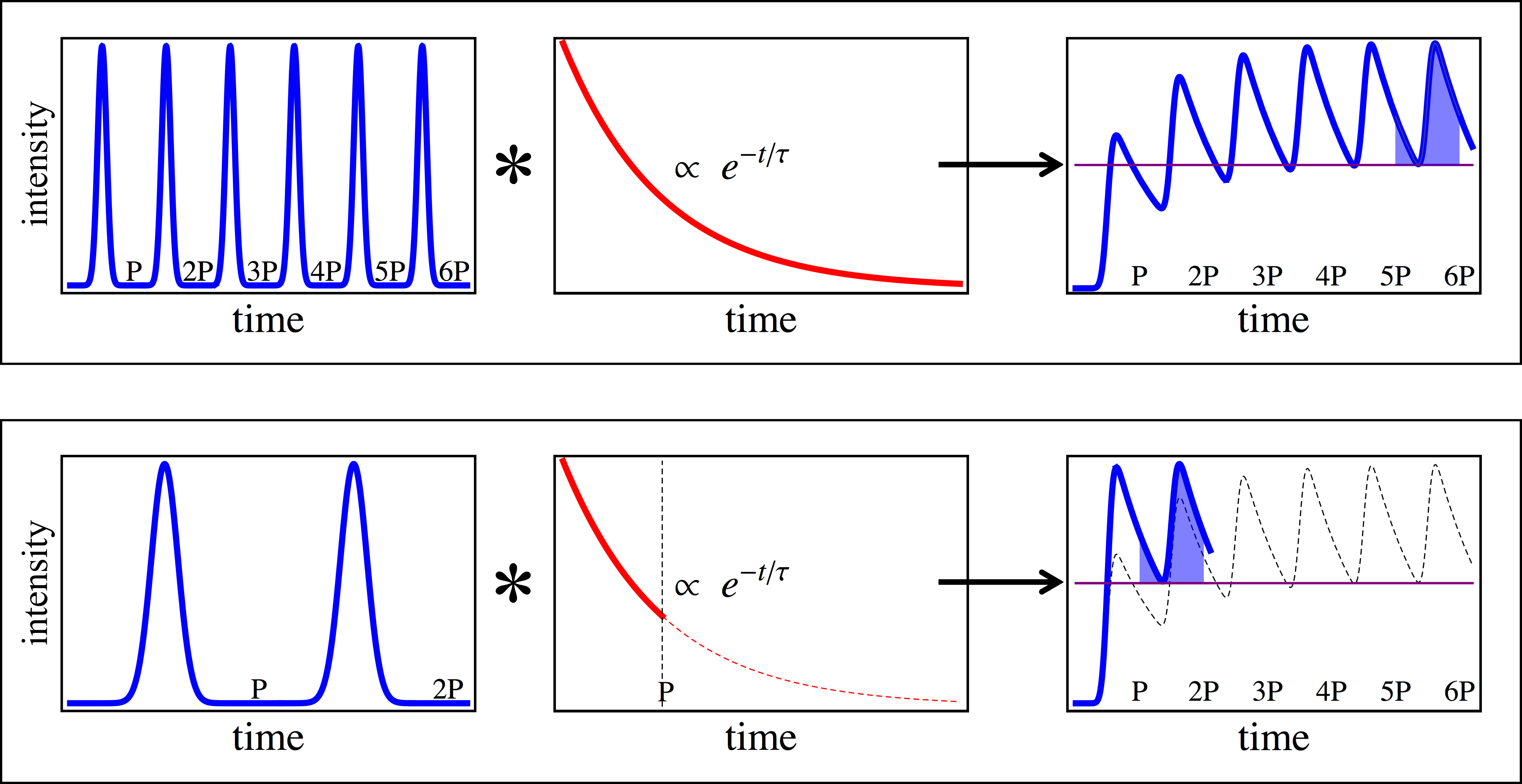}
\caption{A schematic of the different methods for generating noiseless
  scattered profiles, for simulations or forward fitting. The top
  panel shows the \textit{long train} method, and the bottom panel the preferred \textit{train}
  method.}\label{fig:schmodel}
\end{figure*}

The scattering geometry considered in this work is shown in Fig.~\ref{GeoGraph}. 
The radio emission from a pulsar at a distance $D$ from the observer hits a scattering screen at a distance ${D_s}$ (measured from the pulsar).  The photons are scattered by the screen such that those refracted back to the observer travel along paths of different lengths.  The additional path lengths will cause the observed scatter broadening. 

The geometrical relationship between the delay $t$, with respect to the direct line of sight, and the scattered angle $\bsy{a}$ for a single thin screen is given by \mbox{${t} = {D_s}(1-{D_s}/D){|\bsy{a}|^2}/2{c}$} \citep{Cronyn1970}. The scattered angle relates to the observed ray angle $\bsy{\theta}$ by \mbox{$\bsy{\theta}=\bsy{a}({D_s}/D)$}. 

Following the analysis put forth in \citet{CordesLazio2001} we present
general expressions for the probability density function ${f_t}$,
which describes the probability that a given photon from the pulsar
will be delayed by a time $t$, in Sec. \ref{sec:scatscreen}. We
simulate scattering imprints on pulse profiles by convolving an
intrinsic pulse train with a given temporal broadening function $f_t$.
Thereafter, we fit the simulated data with a set of methods and test
how accurately they reproduce the simulated values of $\tau$.

\subsection{Simulated data} \label{sec:simdata}

\subsubsection{Simulated Intrinsic Pulse}\label{sec:simpulse}
The intrinsic pulse profile in our simulated data consists of a single
Gaussian component with a chosen duty cycle and pulse period ($P$).
When necessary, multiple pulse periods are stitched together to form a
pulse train. We investigate a typical slow pulsar with ${P}=1$~s and a
duty cycle of 2.5\% and fast pulsar with ${P}=20$~ms with a duty cycle
of 10\%.

The mean flux spectral index for all the simulations is chosen to be
$\gamma = 1.6$, where the mean flux is ${S_{m}} = \nu^{-\gamma}$. We
use $\gamma$ as the flux spectral index and $\alpha$ as the $\tau$
spectral index to avoid confusion. The mean flux of the pulsar is
normalised such that it has a maximum value of unity for the lowest
studied frequency. The use of a Gaussian template is motivated by the
fact that a large fraction of pulsar profiles are relatively well approximated by
this shape (e.g.
\citealt{KarastergiouJohnston2007,JohnstonWeisberg2006}). The drawback
is that a single Gaussian shape is not representative of all pulsar
profiles, especially not of older, slower pulsars. In Sec.~\ref{sec:geoeffects}
we briefly study possible scattering impacts on a more
complex profile shape based on the average profile of PSR~B1237+25.

\subsubsection{Simulated Scattering Screen}\label{sec:scatscreen}

We consider nearby sources by choosing a standard distance of $D=3.0$~kpc
from the observer to the pulsar with a mid-way scattering screen
at ${{D_s}} = 1.5$~kpc. We assume we can treat the scattering screen
as a cold plasma for which the scattering angle
$\bsy{a} \propto \nu^{-2}$, with $\nu$ the observing frequency
\citep{Lang1971}. For a distribution in scattering angles the standard
deviation will follow the relation
$\sigma_{\bsy{a}} \propto \nu^{-2}$. The scattering strength of the
screen is fixed by choosing a proportionality constant in this
relation. We use $\sigma_{\bsy{a}} = 3$~mas at 1~GHz, such that
$\sigma_\theta = 1$~mas at distances $D= 1$~kpc and ${D_s} =1/3$~kpc,
as suggested in \citet{CordesLazio2001}.

We present two analytic expressions for temporal broadening functions associated with isotropic and anisotropic scattering by an infinite thin screen, and thereafter discuss a ray-tracing experiment by which numerical broadening functions for alternative configurations can be built up.\\

\noindent \normalsize{\textit{Isotropic broadening functions}}\\

\noindent The temporal broadening function for a two dimensional scattering
screen that scatters photons isotropically and can be described by a
circularly symmetric Gaussian probability distribution in the
scattering angle, takes the following form

\begin{align}
{f_{t}} &= \tau^{-1}e^{-t/\tau}{U(t)}\label{eq:ftiso}\\
\tau &= D_{s}'\sigma_{a}^2/{c}\label{eq:tau}\\
{D_{s}'}&= {D_s}(1 - \f{{D_s}}{D}),
\end{align}
where $\sigma_a$ is the standard deviation in scattering
angle ${\bsy{a}}$ in each coordinate direction and ${c}$ is the speed of
light \citep{CordesLazio2001}. The unit step function, ${U(t)}$, ensures that we only consider time $t >0$. Eq.
\eqref{eq:ftiso} is valid for a scattering screen that is infinite transverse to the line of sight. 
Such an infinite screen will conserve flux coming from the pulsar; every
ray scattered out of the line of sight is compensated by another ray
that hits the screen and is scattered back into the line of sight.
For finite screens the total flux of the source may not reach the observer
depending on the physical extent of the screen and the observing
frequency. (See discussion in the Sec. \ref{sec:geoeffects})\\

\noindent \normalsize{\textit{Anisotropic broadening functions}}\label{sec:aniscat}\\

An anisotropic scattering screen would, in general, lead to the observation of elongated images of point sources behind such screens. These effects have, for example, been observed in the imaging of pulsar B0834+06 at 327 MHz \citep{Brisken2010}.

We consider an anisotropic scattering mechanism that can be modelled by an asymmetric Gaussian distribution in scattering angles. The temporal broadening function will now be dependent on two different characteristic scattering times, say $\tau_x$ and $\tau_y$, corresponding to the two dimensions that define the scattering screen, and has the form

\begin{align}
{f_{t}} &= \f{1}{\sqrt{\tau_x\tau_y}}e^{-\f{t}{2}(\f{1}{\tau_x}+\f{1}{\tau_y})}{I}(0,\f{t}{2}(\f{1}{\tau_x}-\f{1}{\tau_y})),
\label{eq:ftani}
\end{align}
where ${I}$ is the modified Bessel function of the first kind. The scattering times are defined as in eq. \eqref{eq:tau} with $\tau_{x,y} =D_{s}'\sigma_{a_{x,y}}^2/{c} $ which reduces to the isotropic case for $\tau_x = \tau_y$. \\

\noindent \normalsize{\textit{Numerical broadening functions}}\label{sec:numbroad}\\

In \citet{CordesLazio2001} it was shown that the broadening function for a disk-shaped isotropic scattering screen centred on the line of sight can be created by amending eq. \eqref{eq:ftiso} with a unit step function, such that the broadening function becomes ${f_t}$ ${U}(t_{max}-t)$. This means that beyond a maximum observation angle $\theta_{max}$, no photons are scattered back into the observer's line of sight. The maximum angle has a geometric path length correspondence to a maximum time delay ${t}_{max}$, as measured with respect to the direct line of sight.

We expand on these effects by means of a ray-tracing code that allows us to place a screen of any size between the source and the observer. We simulate a rectangular screen as a toy model: by specifying a screen's width and height, its position along the line of sight (${D_s}$) and its two-dimensional offset with respect to the line of sight, the scattering geometry is defined. 

The temporal broadening function is constructed by calculating the probability that a given photon hitting the truncated screen will reach the detector. Binning these probabilities according to the associated arrival times for each photon allows us to build up the broadening function with a required time resolution.

\subsubsection{Scattered profiles}\label{sec:scatprof}

Fig.~\ref{fig:schmodel} gives a schematic overview of two methods for
producing scattered profiles. The method shown in the top panel represents the most intuitive way of creating folded scattered pulse profiles: a train of Gaussian pulses are convolved with a broadening function resulting in a train of broadened pulse shapes. A broadened pulse towards the end of the train (shaded) is representative of the average pulse profile. We refer to this method as the \textit{long train} method. 

\begin{figure}
\centering
\includegraphics[width=\columnwidth]{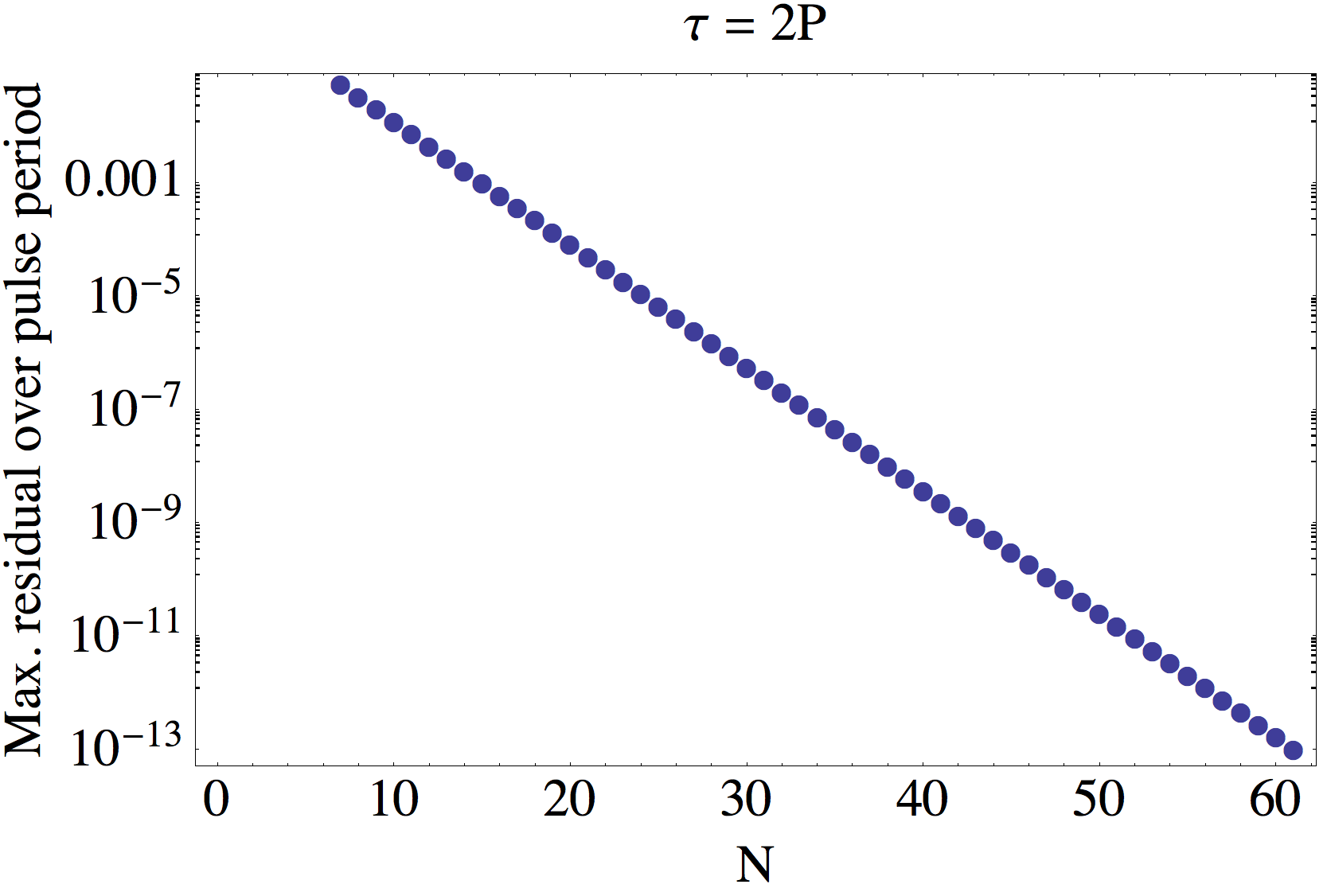}
\caption{The residual differences of the last broadened pulses, as calculated from the analytic expressions for the convolutions in the train method and the long train method respectively, after having corrected for a constant scaling factor equal to the integral of the broadening function up to $t=P$ \mbox{(viz. $1 - e^{-P/\tau}$)}. The residuals are normalised to the area of the pulse and expressed as a function of the number of pulses ($N$) used in the pulse train of the long train method. The broadening function for the long train method is defined for all $t$, whereas for the train method the broadening function is truncated at the pulse period. In this example $P = 0.4$ s and $\tau = 2P$. The residual can be made arbitrarily small for larger scattering values, but will require ever larger $N$ values.}
\label{fig:convcomp}
\end{figure}

An alternative method is shown in the bottom panel of Fig.~\ref{fig:schmodel}. This method, which we simply call the \textit{train} method, convolves a Gaussian pulse train of two pulses with a truncated temporal broadening function defined over a single pulse period (solid line). The second broadened pulse is then extracted as the averaged pulse profile (shaded). The outcome of the long train method (scaled by a constant factor and overplotted in dashed lines in the right-most frame) converges to the same average profile shape after a sufficient period of time. Fig. \ref{fig:convcomp} shows the convergence of the two methods. The residuals between the last broadened pulse resulting from the train method and the long train method are shown to decrease as the pulse train length ($N$) in the latter is increased.

We note that the long train method requires care in the choice in the length of both the
pulse train and the exponential, to ensure that the convolution converges to a constant set of pulses. Typically the required pulse train length increases with an increase in scattering, thereby also increasing the computation time. The advantage of the train method is that it's both computationally less expensive and its accuracy is not dependent on the length of the broadening function or the pulse train (beyond 2 pulses). This motivates us to make use of the \textit{train} method throughout this paper.

Finally, after generating the noiseless scattered profiles, Gaussian
noise is added to simulate an observation with a chosen peak signal to
noise ratio (SNR). Starting from an un-scattered pulse with an
off-pulse baseline of zero, all methods described above will, in the
case of high levels of scattering (large $\tau$ compared to the period
$P$), result in profiles with a raised off-pulse baseline. We use a
Gaussian kernel to smooth the noisy scattered profiles. From the smoothed profile we determine the raised baseline value to subtract. The impact of this on the fitting will be
discussed in Sec.~\ref{sec:end2end}.

The frequency range over which we simulate data is chosen to create
severe scattering imprints at the lowest considered frequency, where
$\tau$ is a substantial fraction or even larger than the pulse period.
In these cases special care must be taken in the forward fitting process, as
discussed the next section. Although not presented in this paper, the
simulations can be extended to include scattering screens with
Kolmogorov (turbulent) scattering statistics.

\subsection{Fitting techniques} \label{sec:fittech}

Forward fitting techniques create models to fit the data by convolving
an estimated intrinsic pulse shape with a parameterised broadening kernel that
describes the response of the ISM \citep{Ramachandran1997}. In general, it is difficult to obtain a model for the intrinsic pulse shape as pulsars often exhibit frequency evolution. In this paper we approximate the intrinsic pulse shape with a Gaussian function, with the caveats already stated.

An example of a forward fitting method is an Exponentially Modified Gaussian (EMG) method. In its most basic form this  gives the analytic expression for a single Gaussian pulse
convolved with a one-sided exponential function. Such a simplistic implementation would not work well for large scattering times compared to the
pulse period.

As mentioned earlier, inverse techniques perform a deconvolution of the
broadened pulse shape through e.g the \textit{CLEAN} algorithm
\citep{Bhat2003CLEAN}. We focus on forward
techniques. As shown in eq. \eqref{eq:tau}, the spectrum of a Gaussian isotropic
scattering model has the form $\tau \propto \nu^{-4}$. We test the
efficiency of fitting methods by determining how accurately they
replicate this theoretical spectrum.

We use the \textit{lmfit} package in python for our fits
\citep{Newville2014}. We use our methods of generating noiseless
scattered profiles, described above, as the fitting functions.The method fitting parameters are the underlying Gaussian intrinsic pulse components, i.e. the width ($\sigma$), mean ($\mu$) and amplitude ($A$), as well as the characteristic scattering time ($\tau$) and a DC offset, described in the next section.


The aim is to fit the simulated data for a range of frequencies to
thereby obtain a spectrum of the characteristic scattering time,
$\tau$. In doing so we bear in mind that all observed scattered
broadened profiles will typically result from integrating over some
frequency band. The $\tau$ value obtained for fitting a monochromatic
simulated scattered profile at frequency ${f_m}$ corresponds to the
$\tau$ value associated with the central observing frequency ${f_c}$
and a bandwidth $\delta {f}$ by
\begin{equation}
{f_m} = 10^{[\log_{10}({f_c} + \delta {f}/2) + \log_{10}({f_c} - \delta {f}/2)]/2}
\label{eq:fcfm}
\end{equation} 
or inversely,
\begin{equation}
{f_c} = \frac{1}{2}\sqrt{\delta {f}^2 + 4{f_m}^2}.
\label{eq:fmfc}
\end{equation}
This equation provides the relevant frequency for a $\tau$ measurement
given any $f_c$ and $\delta f$, assuming a power law spectrum.
Therefore computed $\tau$ values should be plotted against calculated
${f_m}$ values to obtain meaningful fits. We will revisit the impact of this in the next section.
We note that in our methodology we fit all frequency bands independently, whereas more constraints could be placed by simultaneously fitting for $\tau$ and the pulse parameters across a broad bandwidth, if that were available. 

\begin{figure*}
\centering
\subfigure{\includegraphics[width = \columnwidth]{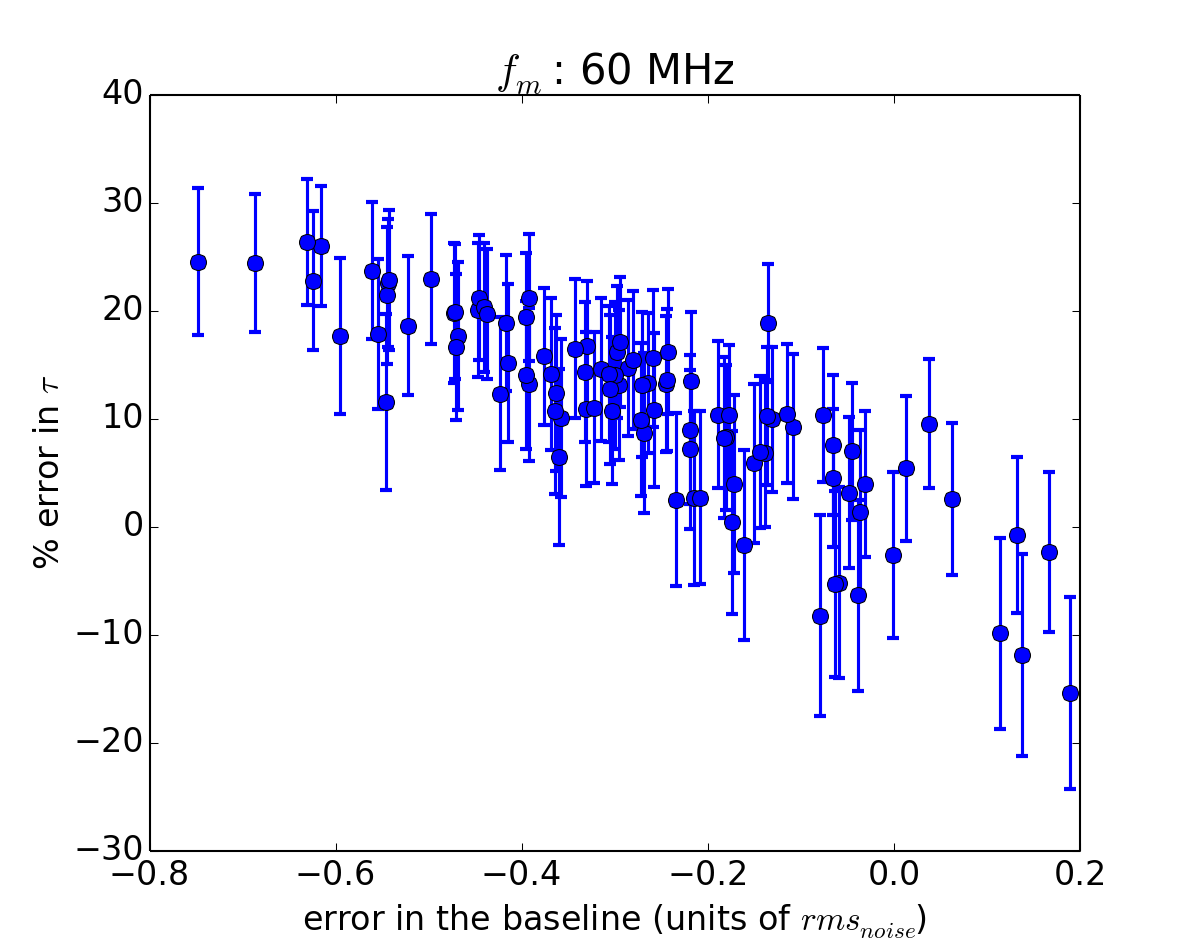}}
\subfigure{\includegraphics[width = \columnwidth]{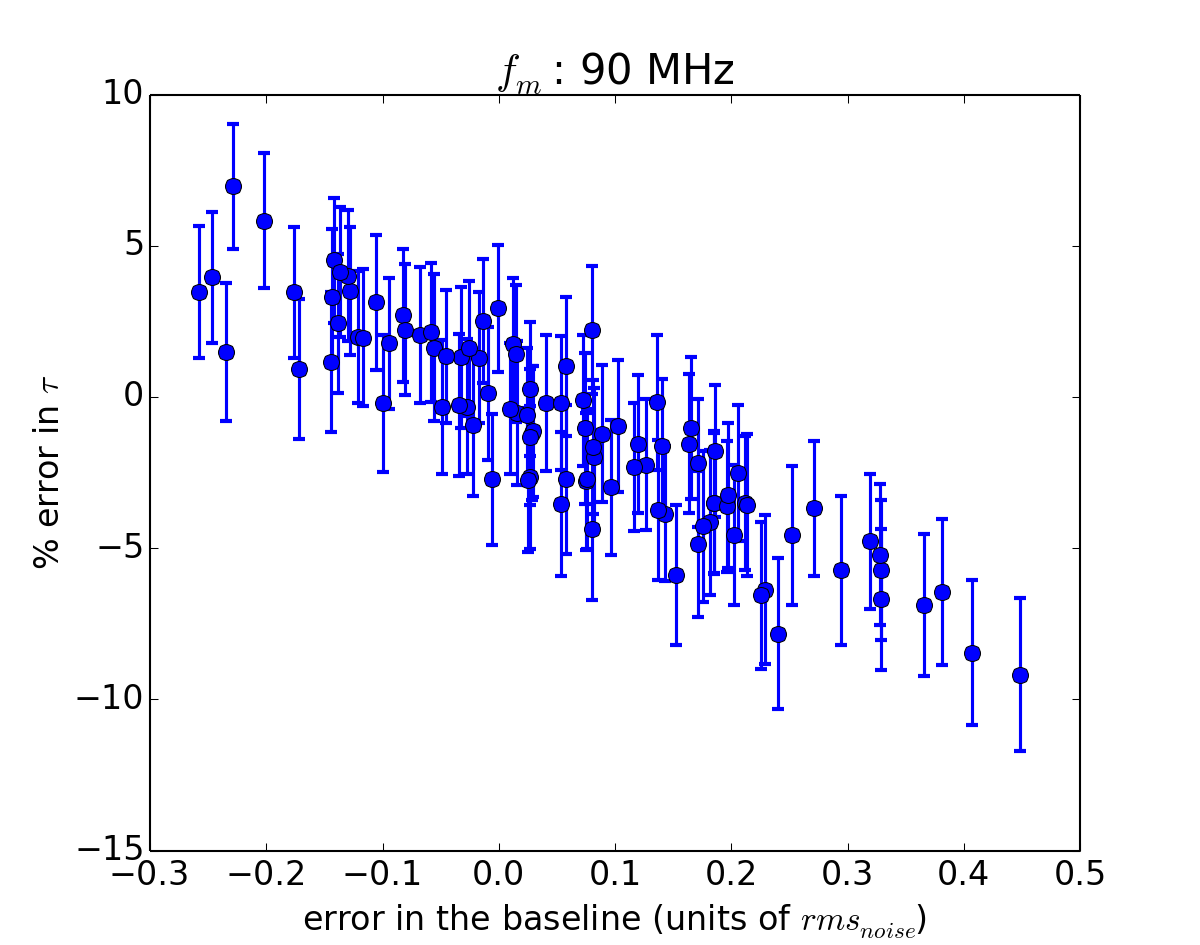}}
\caption{Uncertainties in the obtained $\tau$ values as a function of the error in estimating the baseline of the scattered profile, shown for two distinct frequency values. Each panel represents 100 independent realisations of Gaussian noise. The data are simulated as described in the text. }
\label{fig:basestats}
\end{figure*}

\begin{figure*}
\centering
\includegraphics[width = \textwidth]{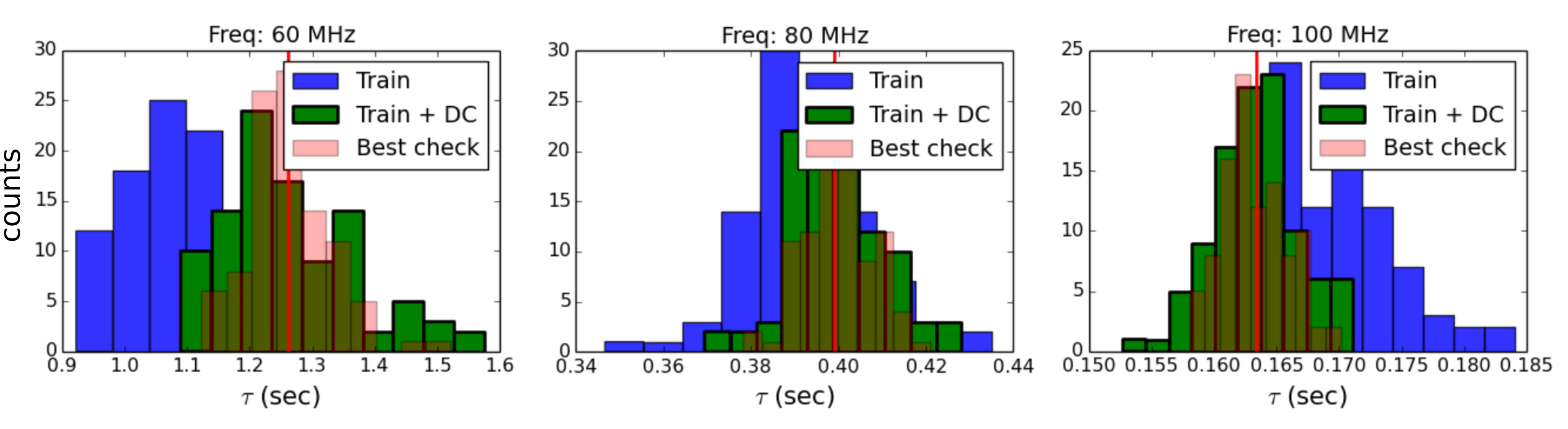}
\caption{Histograms showing the obtained spreads in $\tau$ values at three sample frequencies, using different methods. The red line indicates the true value of $\tau$. The \textit{best check} histogram shows the distribution obtained from fits where the noiseless minimum of the profile is known exactly, and is therefore an idealised reference case. The \textit{train} and \textit{train + DC} methods show the spreads in $\tau$ using the train method with our without fitting for a DC-offset. In the case where the method does not fit for a DC-offset, a Gaussian smoothing kernel is used to estimate and subtract the raised off-pulse baselines.}
\label{fig:hist}
\end{figure*}
\section{Results}\label{sec:results}

\subsection{A simple end-to-end experiment}\label{sec:end2end}

\begin{figure}
\centering
\includegraphics[width = \columnwidth]{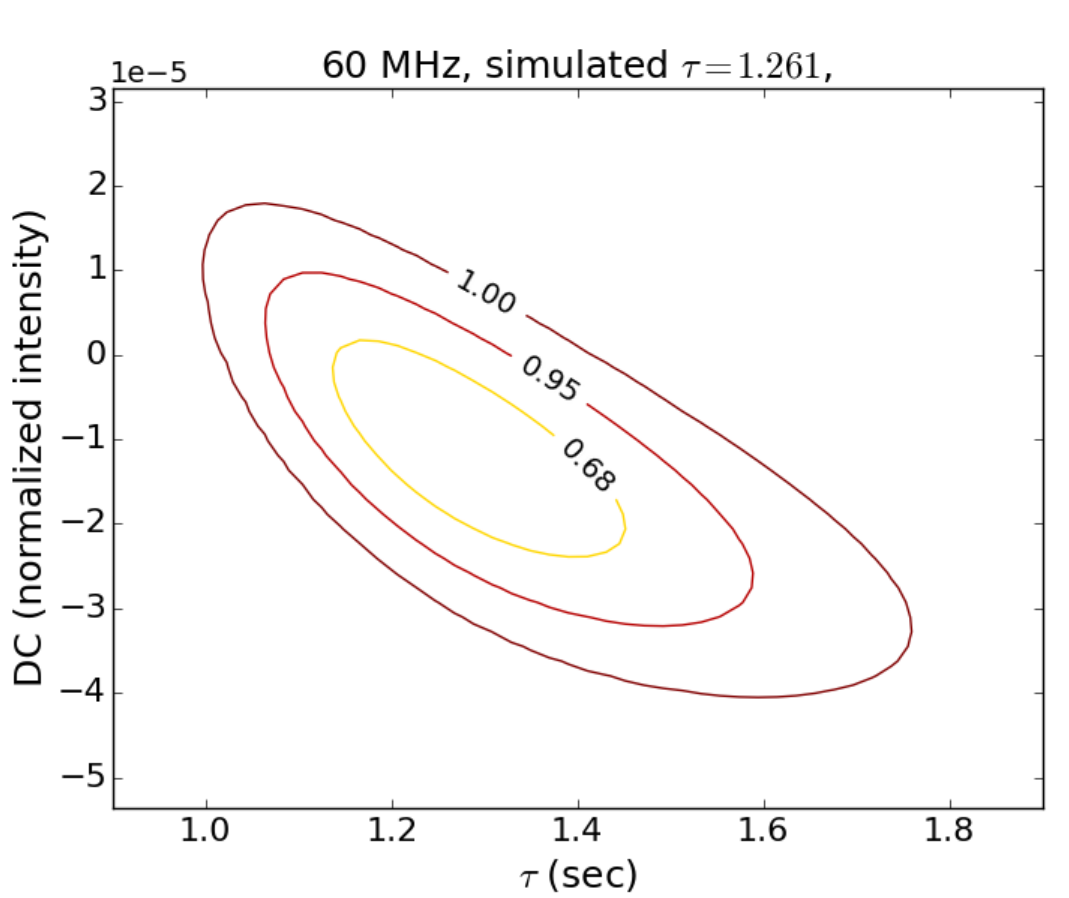}
\caption{The confidence intervals (1$\sigma$ - 3$\sigma$) associated with the obtained fits for the vertical offset parameter, $DC$, and characteristic scattering time, $\tau$.}
\label{fig:taudc}
\end{figure}

\begin{figure}
\includegraphics[width = \columnwidth]{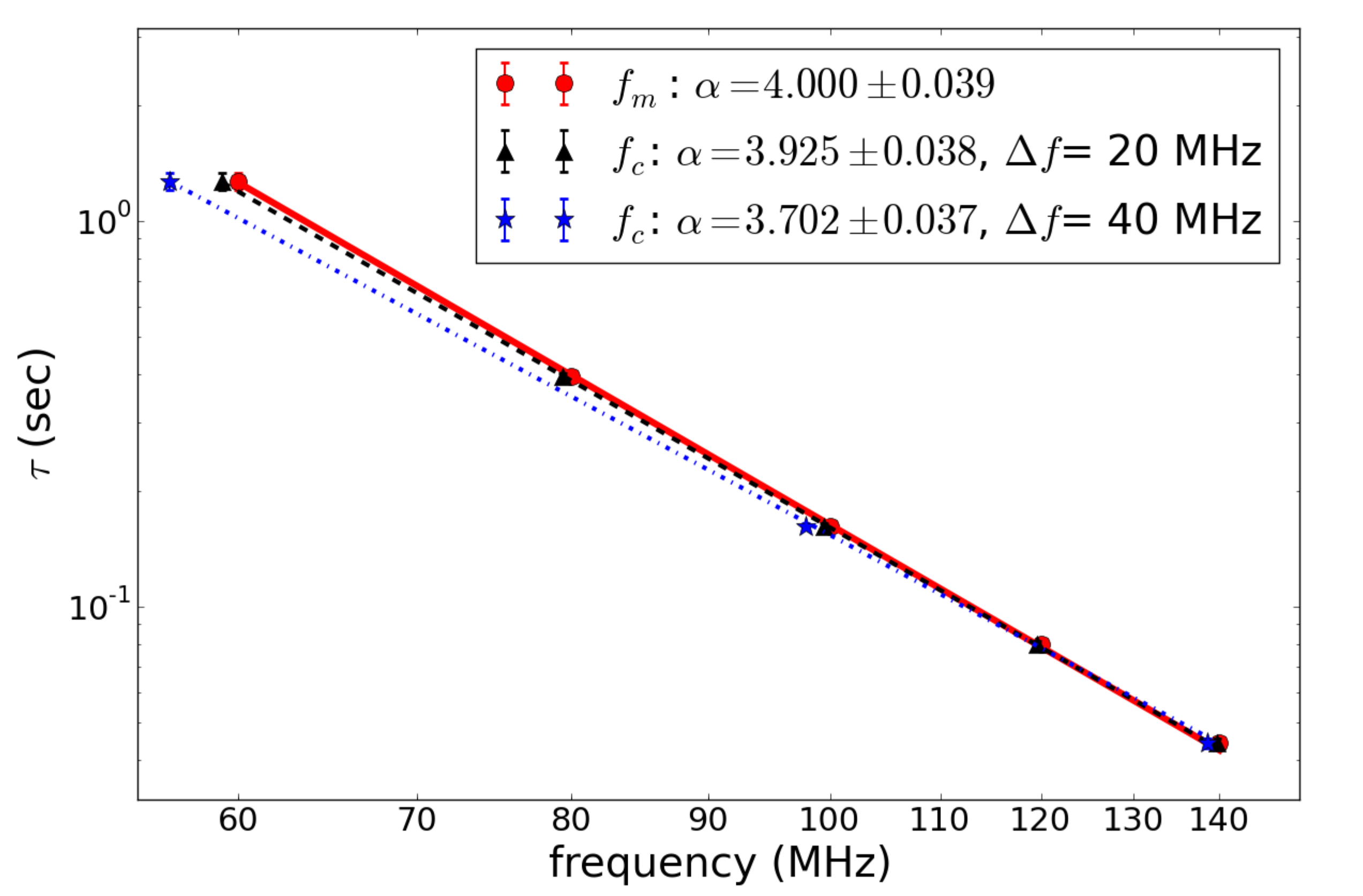}
\caption{Deviations from the theoretical characteristic $\tau$ spectrum are observed as the bandwidth of the simulated data increases. The obtained $\tau$ associated with a central frequency ${f_c}$ and bandwidth $\delta f$ relate to the simulated result produced by a monochromatic frequency input ${f_m}$ through eq. \eqref{eq:fcfm}. }
\label{fig:fcfm}
\end{figure}

Before moving onto detailed results, we describe a simple end-to-end
experiment. We simulate an intrinsic pulse shape scattered by an
isotropic infinite scattering screen as described in Sec.
\ref{sec:simdata}. The frequency range over which we conduct the
simulations is 60 - 140~MHz. Within this frequency range the pulse
profiles of the pulsar with $P = 1.0$~s and a duty cycle of 2.5~\%, are notably scattered and in the
severest of cases (lowest frequencies) the scattering tail wraps
around.

Initially, we generate profiles monochromatically at 60 and 90~MHz 
with a peak SNR of 20. We fit the noisy profiles using the train method. 
We find that the determination of the off-pulse baseline of the noisy data has an
important effect on the fit, which we investigate as follows. As
mentioned, we use a Gaussian kernel to smooth the simulated data
and from this smoothing function estimate the off-pulse baseline, 
denoted in the following as $b$.
The effect of the error in accurately determining the baseline, on the
subsequently fitted $\tau$ value, is shown in
Fig.~\ref{fig:basestats}. Here, we compare the fitted $\tau$ values to
the outcome when the off-pulse baseline is known exactly. The plots 
show a percentage error in $\tau$,
$100(\tau_{known} - \tau_{fit})/ \tau_{known}$ plotted against the
error in the determined baseline in units of the noise,
($b_{known} - b_{est.})/rms_{noise}$, for 100 realisations of the
noise at 60 and 90~MHz.

The error in the $\tau$ value increases systematically as the error in
the baseline value grows, and is greater at the lower of the two
frequencies. At 60 MHz, an error of 10 - 20\% in the $\tau$ value
is already possible for baseline errors below $1 \sigma$ of the
noise. As the frequency increases, the error in $\tau$ for a given
error in the baseline drops significantly. We also note that at lower frequencies 
the $\tau$ value is more likely to be underestimated, 
whereas at higher frequencies $\tau$ is more likely to be overestimated.

This result prompted us to investigate an improved model. In Fig.~\ref{fig:hist} we show the impact of adding a DC-offset to the parameter set for which the model fits. In the \textit{train~+~DC} model we therefore no longer estimate the off-pulse baseline in order to subtract it from the profile before fitting the data, but we allow the model to fit for a best off-pulse baseline value.
The spread in $\tau$ values in Fig.~\ref{fig:hist} represents 100 noise realisations, 
and the data are simulated as for Fig.~\ref{fig:basestats}. Introducing a DC offset to the train method, ensures that the estimated $\tau$ values are centred on the true value (red line). The histogram labelled \textit{best check} gives the spread in obtained $\tau$ values when the noiseless off-pulse baseline of the simulated data is known exactly. We note that the \textit{train~+~DC} model approximates this idealised distribution well. However, at both the low and high end of the frequency range shown, the \textit{train + DC} distributions are more asymmetric than this reference case.

Fig.~\ref{fig:taudc} shows the correlation between $DC$ and $\tau$ and the associated confidence intervals, at 60 MHz. The parameters are anti-correlated, with a correlation coefficient of \mbox{C($DC$,$\tau$) = -0.73}.  This correlation allows an improved accuracy of the obtained $\tau$ value, when including $DC$ as a fitting parameter. 

For the remainder of this paper we will mostly 
make use of the train method with a DC offset. 
The obtained $\tau$ values are therefore subject to uncertainties as inferred from the distributions in Fig.~\ref{fig:hist}. We consider this to best match real observational data, where the baseline determination will always be affected by instrumental noise.

In this end-to-end experiment we also investigate the typical impact
of the bandwidth of an observation on the determined $\tau$ spectrum.
As shown in eq. \eqref{eq:fcfm} the $\tau$ value associated with a fit
of an integrated profile must be remapped to the corresponding
frequency ${f_m}$, before the spectrum is computed.

Fig.~\ref{fig:fcfm} shows the impact on the $\tau$ spectrum fit when
the central frequency values ${f_c}$ are used instead of ${f_m}$.
Simulating the data monochromatically at 20~MHz intervals from 60 to
140~MHz leads to a measured spectral index of \mbox{$\alpha = 4.00 \pm 0.04$},
reproducing the theoretical spectrum well. In contrast a bandwidth of
20~MHz at 20~MHz intervals leads to a spectral index of
\mbox{$\alpha = 3.93 \pm 0.04$}, when plotted against ${f_c}$. An even larger
bandwidth of 40~MHz gives \mbox{$\alpha = 3.70 \pm 0.04$}. To produce this
figure, data were fitted with the train method and the baseline of the
simulated data subtracted before adding noise - to eliminate the
effects of baseline errors as mentioned above.

\subsection{Results for Infinite Screens} \label{sec:resinf}

We now move on to infinite scattering screen setups. We first present
the obtained $\tau$ spectra, followed by the flux spectra. In each
case we consider the two scattering mechanisms discussed in Sec.
\ref{sec:scatscreen}, which serve as examples of either an isotropic
or anisotropic scattering process.

\subsubsection{Tau Spectra}\label{sec:tau}
\paragraph{Isotropic Scattering Tau-spectra}\label{sec:isotau}

\begin{figure*}
\includegraphics[width = \textwidth]{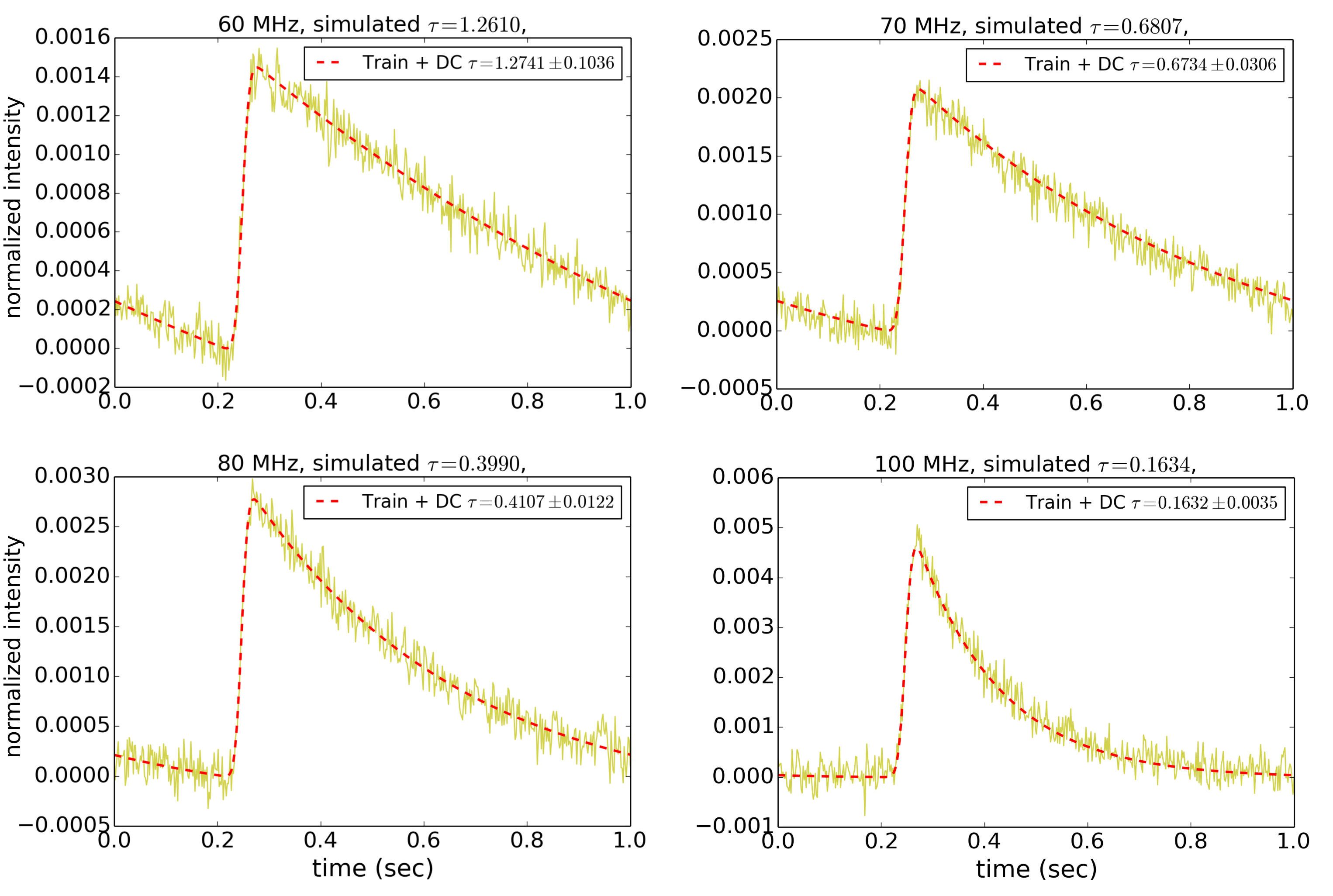}
\caption{Best-fits as produced by the train + DC method for simulated scattered pulse profiles of a ${P}= 1.0$ s pulsar. Fits without the DC-offset parameter (i.e the train method in which the off-pulse baseline is determined using a smoothing kernel) look similar to the DC fits by eye, but produces skewed values for $\tau$, as was shown in Fig.~\ref{fig:hist}. The impacts of these offsets on the power law fit to the $\tau$ spectra can be seen in Table~\ref{table:one}. Profiles are plotted normalised to the area under the pulse - such that at the lowest frequency (viz. 60MHz) in the absence of a non-zero off-pulse baseline, the area under the pulse is equal to one.}
\label{fig:profilesmodelslow}
\end{figure*}

\begin{figure*}
\centering
\includegraphics[width = \textwidth]{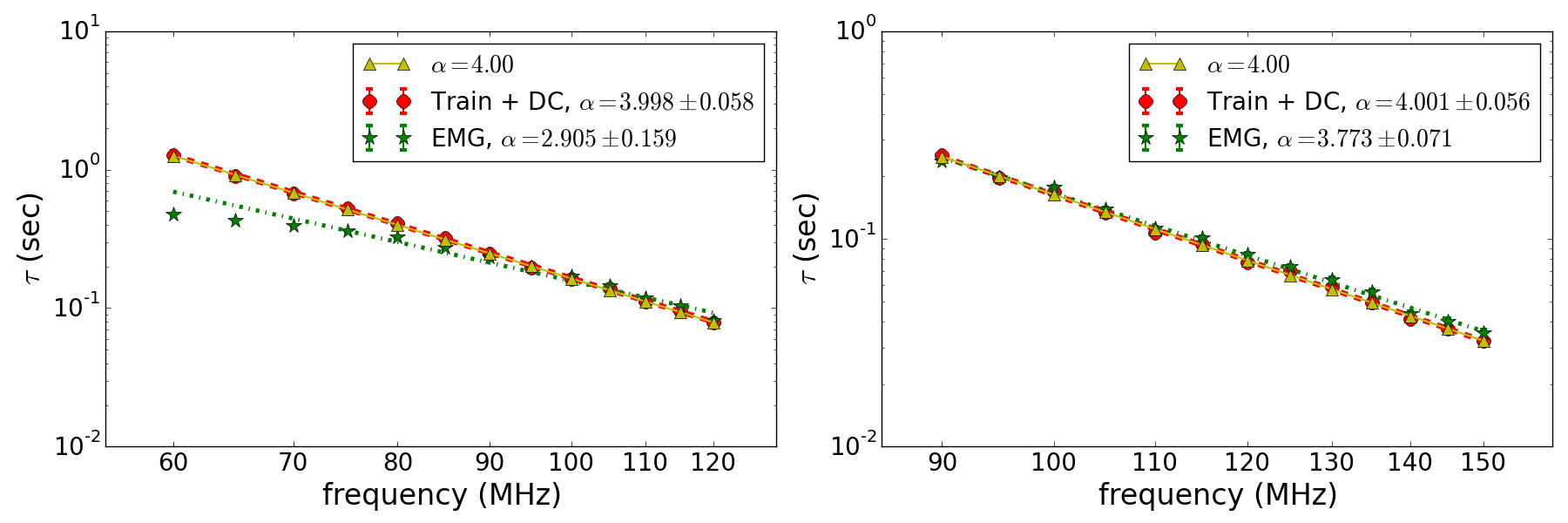}
\caption{The $\tau$ spectra for the 1.0 s pulsar as obtained by the train + DC method. The left-hand panel shows the spectrum at lower frequency values (high scattering), and the right-hand panel at higher frequency values. Even at the levels of scattering associated with the left hand panel, the train + DC method is able to accurately reproduce the theoretical $\tau$ spectrum with a spectral index close to 4. The obtained 1$\sigma$ error bars are of the same order as the marker size and therefore not clearly visible.  The EMG method is included as an example of a method that does not model wrap around scattering tails and can therefore not accurately reproduce spectral index values in data where high levels of scattering are present. Table \ref{table:one} gives additional information about the standard deviations in the $\alpha$ values when rerunning the experiment multiple times and fitting with various methods.}
\label{fig:tauspectra}
\end{figure*}

\begin{figure*}
\includegraphics[width = \textwidth]{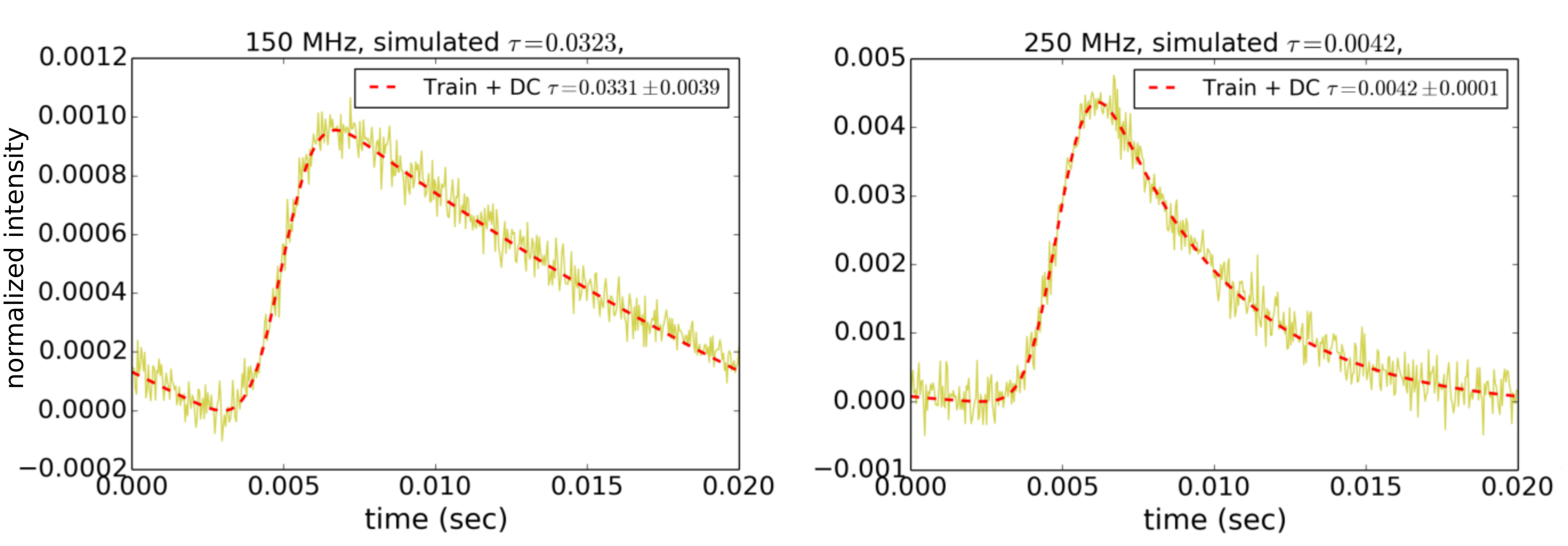}
\caption{Similar to Fig.~\ref{fig:profilesmodelslow}, but for a pulsar with period ${P}= 20$ ms and a duty cycle of 10\%.}
\label{fig:profilesmodelfast}
\end{figure*}

\begin{figure*}
\includegraphics[width = \textwidth]{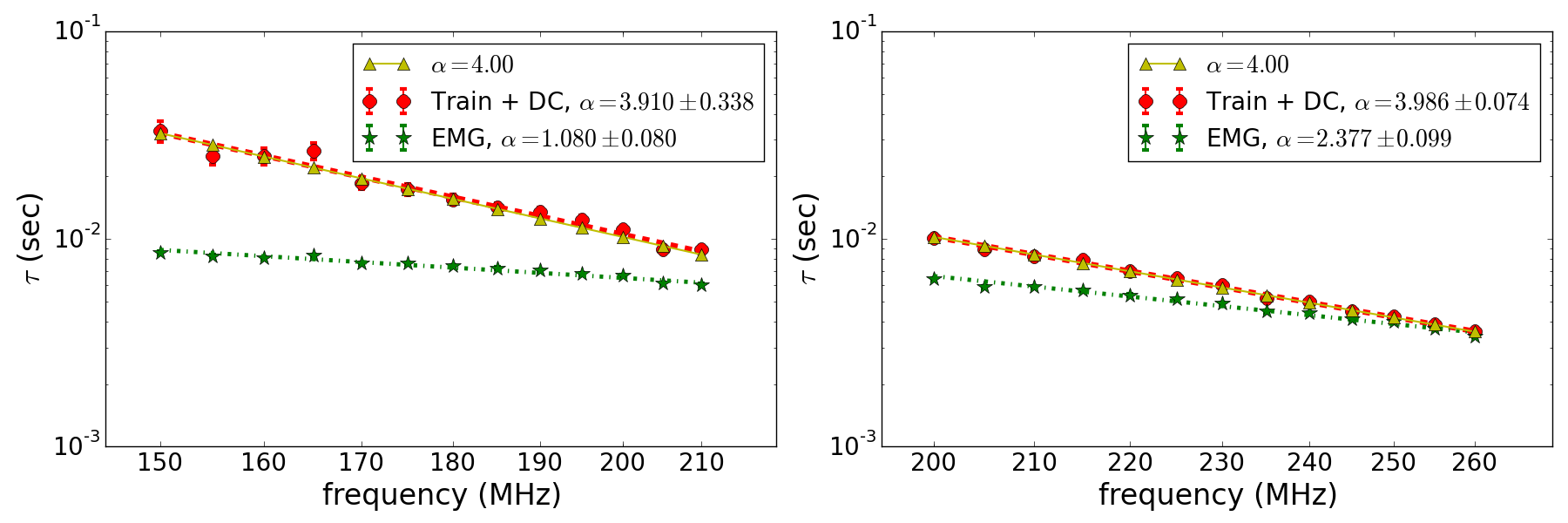}
\caption{The $\tau$ spectra of the modelled millisecond pulsar. The uncertainty in the spectral index obtained by the train + DC method grows as the frequency range is lowered. At these lower frequencies the EMG method deviates most severely from the theoretical spectrum, and continues to underestimate the value of $\alpha$ even for the higher frequency range. The frequency at which the choice of method matters is much higher for the simulated millisecond pulsar than for the slow pulsar in Fig.~\ref{fig:tauspectra}.}
\label{fig:tauspectrafast}
\end{figure*}

\begin{table}
\centering
\caption{The mean spectral indices and standard deviations as obtained from 1000 $\tau$ and subsequent spectra fits. Two pulsars are investigated: a slow pulsar and a millisecond pulsar. For each pulsar the spectral index is determined over two sets of frequencies. The fits and their associated standard deviations improve with a decrease (increase) in scattering (frequency). The different methods are as described in the text.}
\includegraphics[width = \columnwidth]{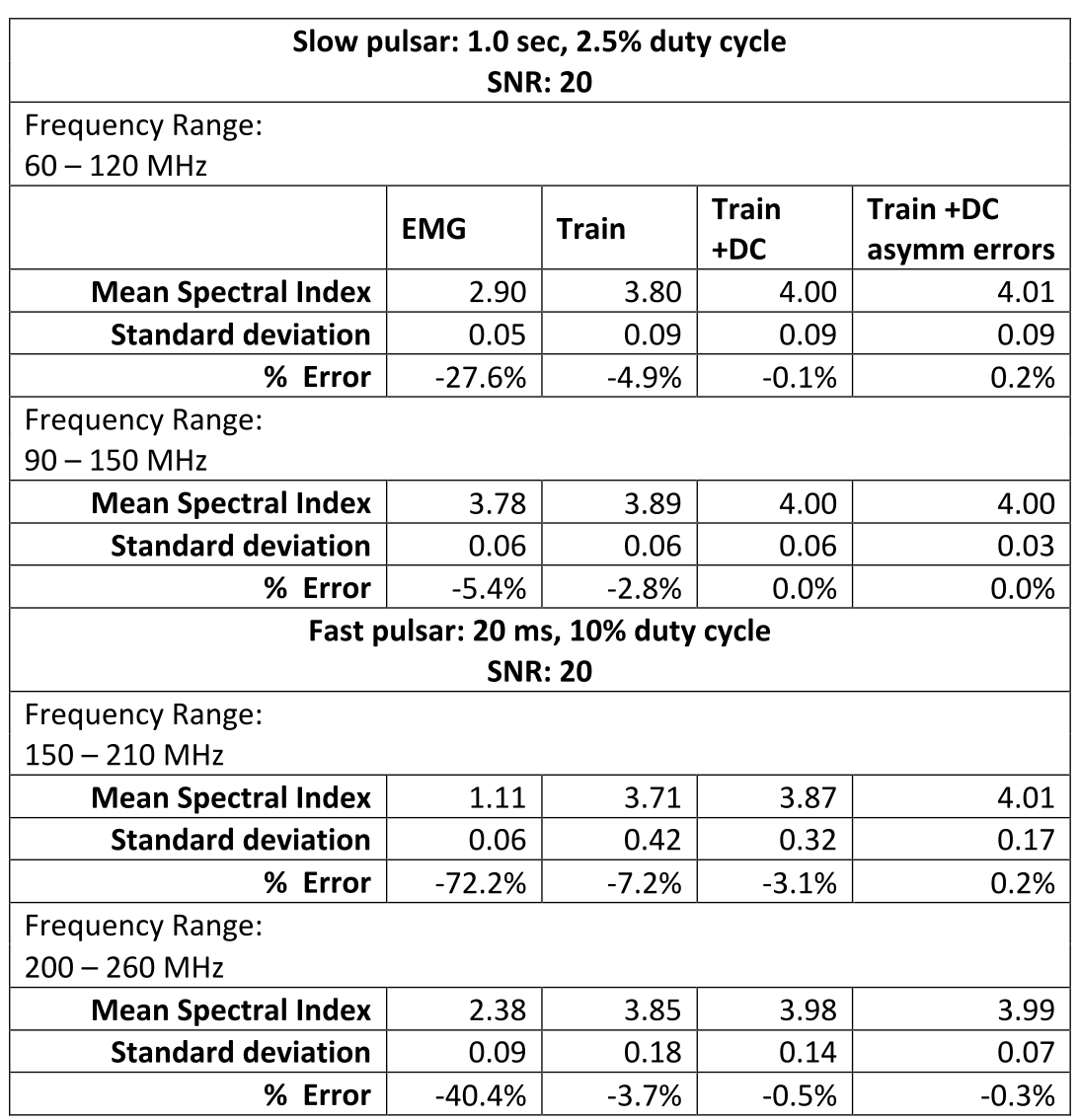}
\label{table:one}
\end{table}

\begin{figure}
\includegraphics[width = 1.1\columnwidth]{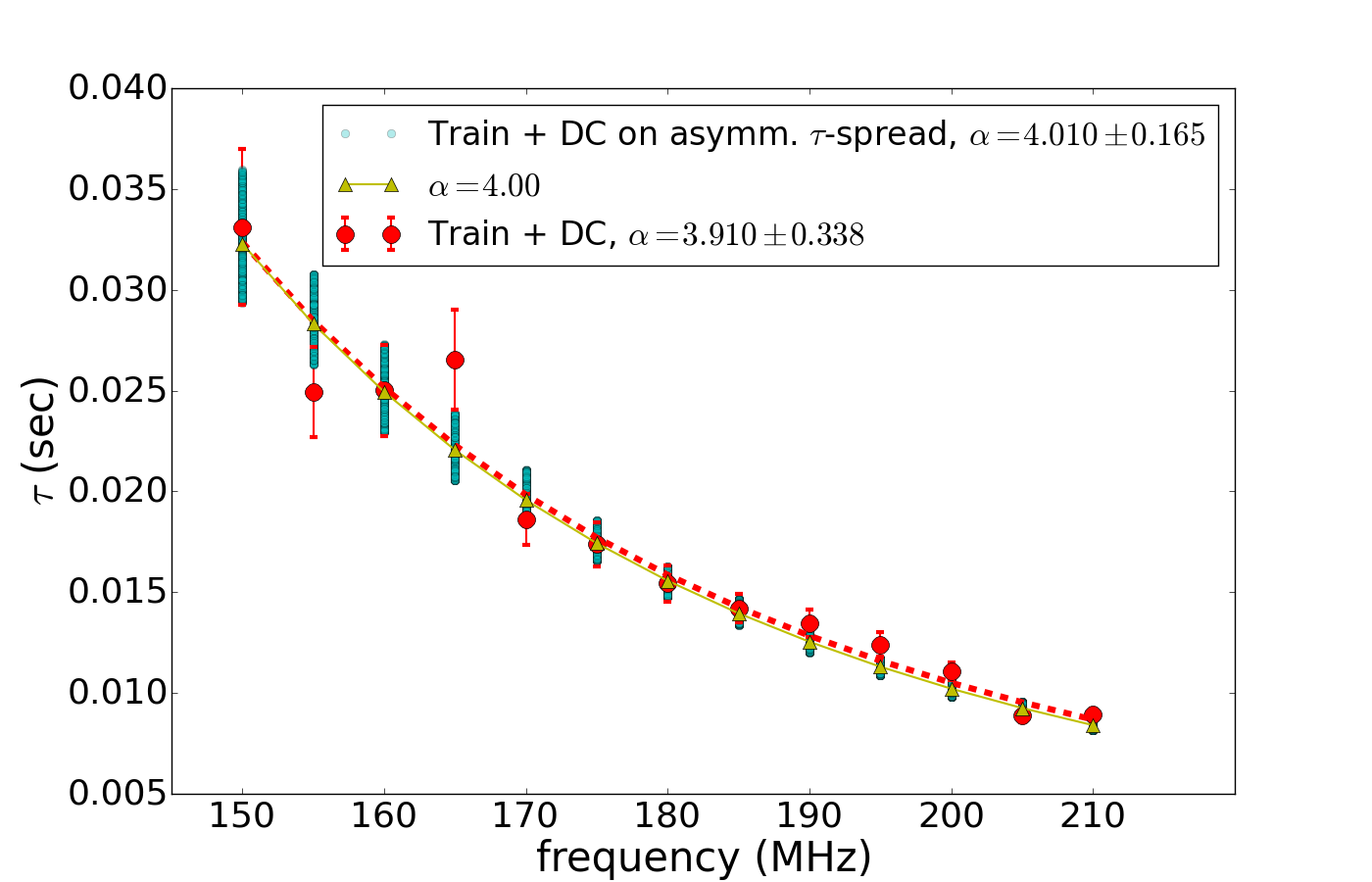}
\caption{The power law fit of the millisecond pulsar is improved when we take the skewness of the $\tau$ distribution (resulting from 1000 noise realisations and subsequent fits) into account at each frequency. The $\tau$ values that fall within the 1$\sigma$ uncertainty of this distribution are shown as cyan circles at each frequency. The train + DC fit (red) is an example of a single noise realisation and subsequent fit. Data are plotted on a linear scale for the $\tau$ spread values to be visible.}
\label{fig:tauspread}
\end{figure}

Isotropically scatter broadened profiles of the two simulated pulsars
described in Sec.~\ref{sec:simpulse} are shown in
Figs.~\ref{fig:profilesmodelslow} (1.0 s pulsar) and
\ref{fig:profilesmodelfast} (20 ms pulsar). The simulated data, as
seen in yellow, again have peak \mbox{$\rm{SNR} = 20$}. The scattering
setup is as described in Sec.~\ref{sec:scatscreen}. A range of
observing frequencies is chosen such that the scattering tail wraps
around at the low end and a typical exponential scattering tail is
observed at the high end.

The profiles are fitted with the train + DC method as described in the previous section.  Fits without the DC-offset parameter (i.e the train method in which the off-pulse baseline is determined using a smoothing kernel) look very similar to the DC fits by eye, but produce skewed values for $\tau$ as was shown in Fig.~\ref{fig:hist}. The impacts of these offsets on the power law fit to the $\tau$ spectra can be seen in Table~\ref{table:one}.

As described in Sec. \ref{sec:scatprof} the train and the long-train methods, with added DC offsets, can perform equally well for all the simulations. A long train method with a too short pulse train will lead to an underestimation of $\tau$.

In the case of severely scattered profiles it is vital to use an accurate fitting method as
the outcome becomes increasingly sensitive to the model parameters. The train + DC method presents the simplest (and 
computationally fastest) way to obtain accurate $\tau$ values.

The $\tau$ spectra for the simulated pulsars are shown over two
frequency ranges in Figs.~\ref{fig:tauspectra} and
\ref{fig:tauspectrafast}. As the frequency increases, the obtained spectral
index value more accurately approaches the simulated index of
\mbox{$\alpha = 4$}. Here we also include a basic implementation of the EMG model not accounting for the pulse period $P$. This method
underestimates $\tau$ significantly at high levels of scattering. At higher frequencies where the scattering tail no longer wraps
around, the choice of method becomes less important. 

We generate 1000 datasets, of simulated scattered profiles
at intervals of 5~MHz within our frequency ranges, per pulsar and run the fits
for $\tau$ and $\alpha$. The outcomes of this process are summarised
in Table~\ref{table:one}. 

For each realisation of the experiment, the train + DC method performs best, and in the case 
of the millisecond pulsar (for which the scattering is more severe overall) has a
lower standard deviation in its estimation of $\alpha$, than the
standard train method. The average
spectral index is calculated over two sets of frequencies for both the
slow and millisecond pulsar. At higher frequencies (lower scattering)
$\alpha$ approaches the simulated value of 4 more accurately. This can for example be seen in the tabulated error values, which represent the error of the mean spectral index with respect to the true value of 4 expressed as a percentage.

In Fig.~\ref{fig:hist} we showed the spread of $\tau$ values
obtained by the train methods and by a \textit{best check} reference case.
As noted even the train~+~DC model
shows a skewed distribution of $\tau$ - especially at low frequencies. 
To take this skewness into account, we update the power law fit as follows. 
From a distribution of $\tau$ values (obtained from several noise realisations at a given SNR level, and subsequent profile fits with the train~+~DC method), we subtract $16\%$ of the values from both the lower and upper end of the distribution to be left with the 1$\sigma$ ($68\%$) central distribution. We now use the values within this range along with their corresponding frequency values to update the $\tau$ spectrum, as shown in Fig. \ref{fig:tauspread}. As can be seen from Table~\ref{table:one} under the column heading \textit{Train + DC, asymm errors}, this process mainly provides a small improvement at extreme levels of scattering, such as for the millisecond pulsar at 150-210~MHz.  For the most part of this article we will therefore not show these improvements.

As is expected the success of the $\tau$ spectra fits and the associated uncertainties are also impacted by the peak SNR value, and will improve as the SNR increases. In Table \ref{table:snr1050} we show the achieved spectral index values after repeating the experiment above at SNR values of 10 and 50. At a SNR value of 50 the theoretical spectrum is best reproduced and the standard deviation in the spread of individual measurements has decreased significantly.

\begin{table}
\centering
\caption{Spectral indices similarly obtained as for Table \ref{table:one}. The simulated data have SNR values of 10 and 50, and show the expected improvements in fits as the SNR values increases. The tabulated values represent a mean $\alpha$ and associated standard deviation after 1000 executions of the experiment.}
\includegraphics[width = \columnwidth]{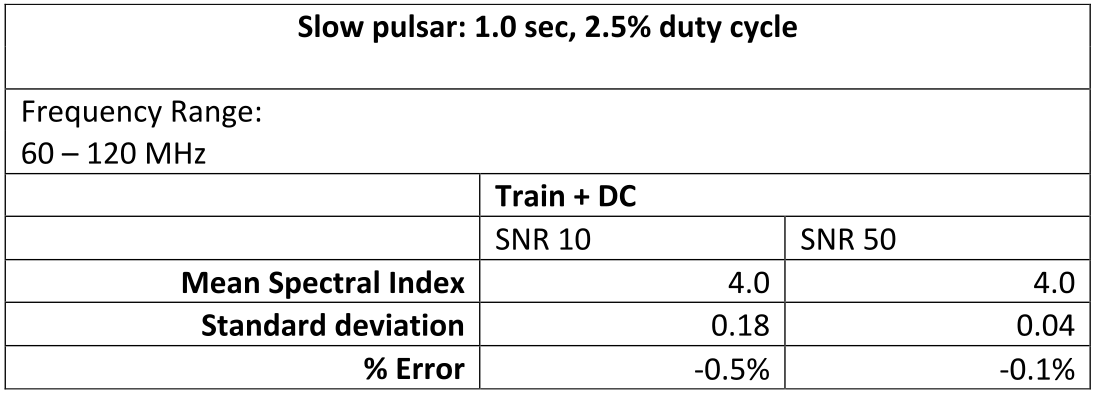}
\label{table:snr1050}
\end{table}

\paragraph{Anisotropic Scattering Tau-spectra}

\begin{figure*}
\centering
\includegraphics[width = \textwidth]{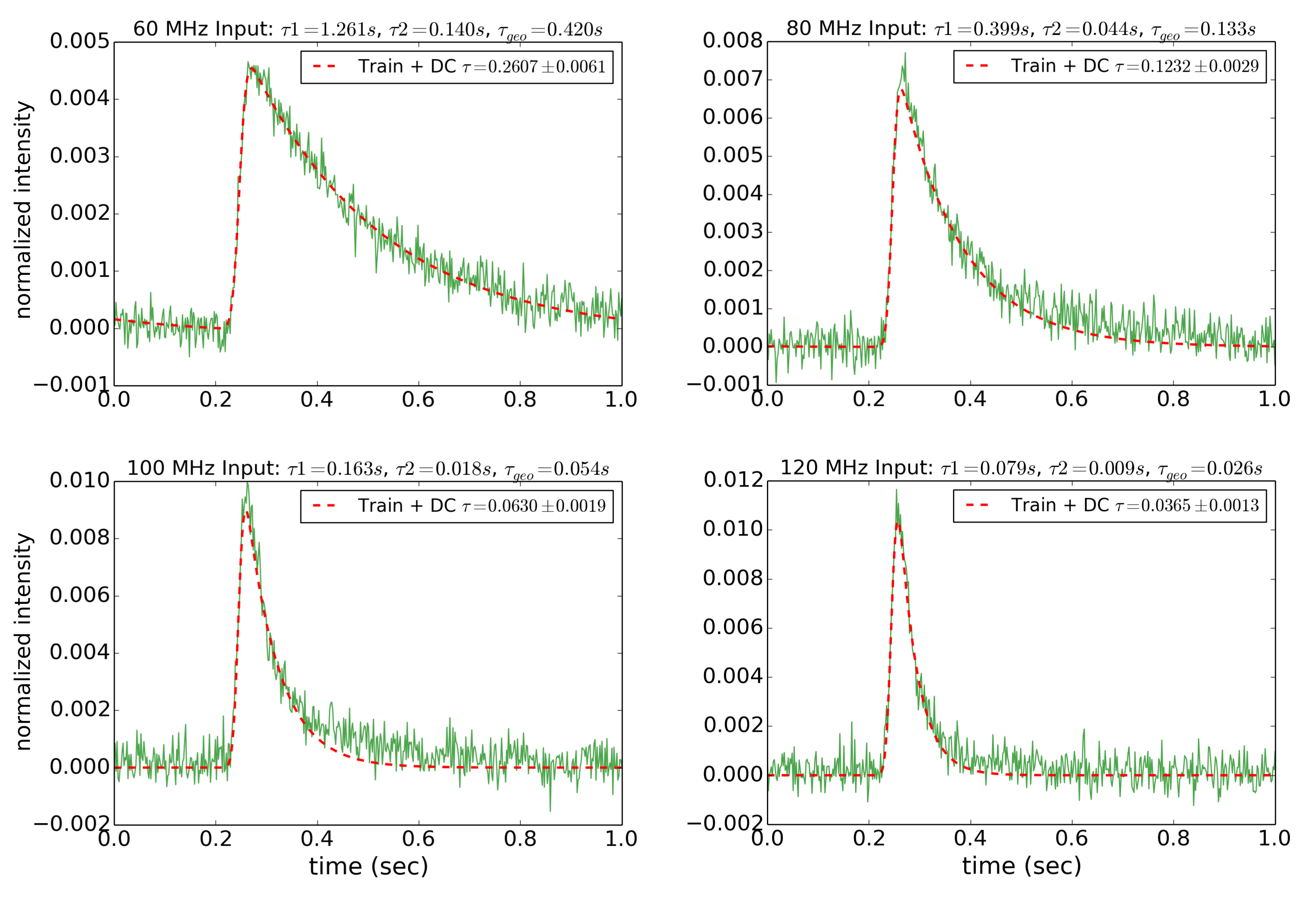}
\caption{Profile fits as obtained by applying the isotropic train + DC method to anisotropically simulated pulse profiles of a $P = 1.0$~s pulsar. The anisotropy is created by choosing a thin screen that scatters 3 times more strongly in one dimension than in the other. At higher frequencies the isotropic fit to anisotropic data leads to a rough estimation of the geometric mean scattering time ($\tau_{geo}$). At low frequencies $\tau_{geo}$ is significantly underestimated. Profiles are plotted normalised to the area under the pulse, such that at the lowest frequency (viz. 60MHz) in the absence of a non-zero off-pulse baseline, the area under the pulse is equal to one.}
\label{fig:profani}
\end{figure*}

\begin{figure}
\centering
\includegraphics[width = 1.1\columnwidth]{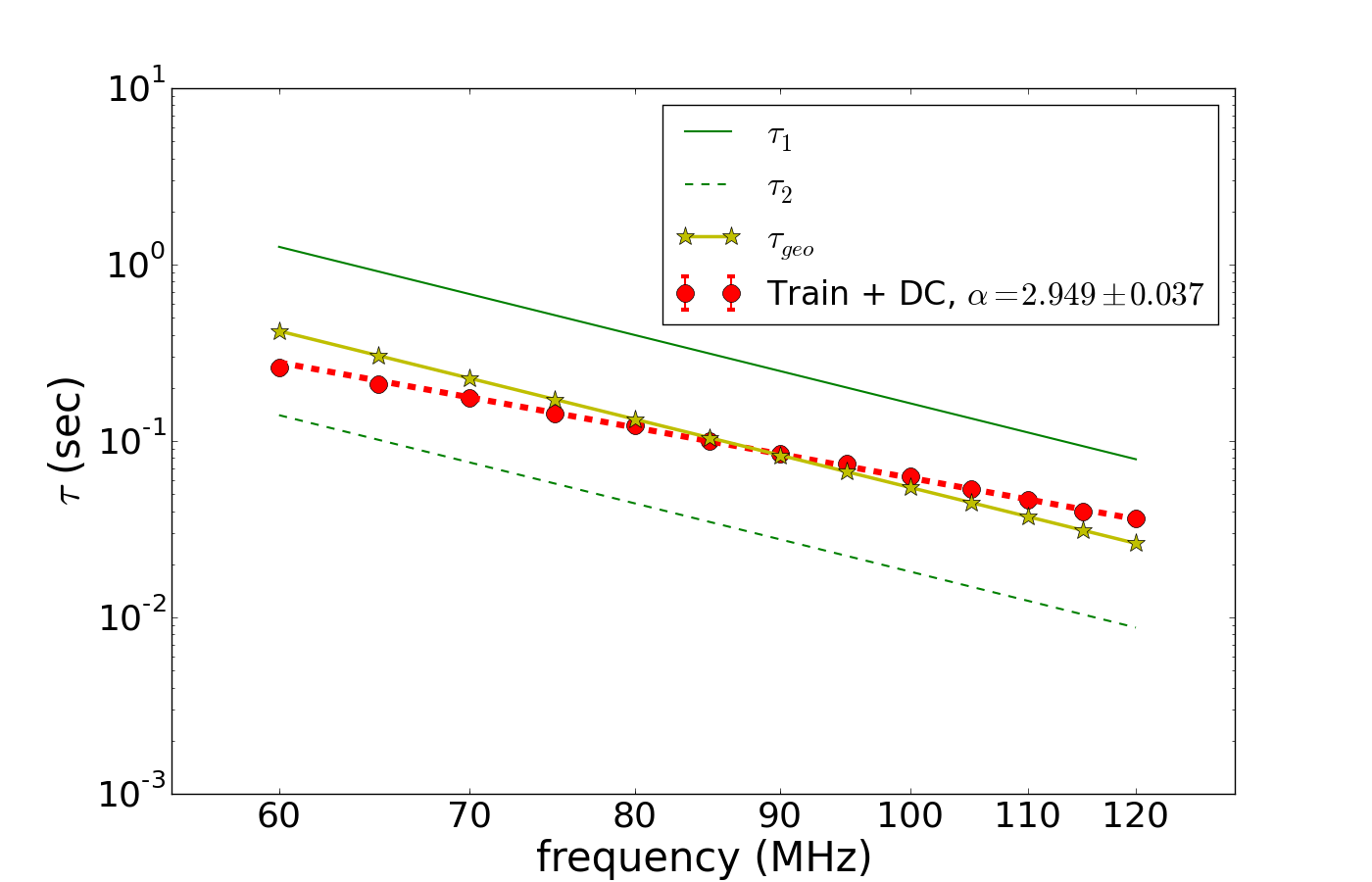}
\caption{The $\tau$ spectrum for an anisotropic scattering event fitted with a single $\tau$ power law is shown in red (thick dashed line, circles). The green lines (solid and dotted without markers) show the simulated values for $\tau_1$ and $\tau_2$. The geometric mean for each of these values, $\tau_{geo} = \sqrt{\tau_1 \tau_2}$ is shown in yellow (stars). The $\tau_1$, $\tau_2$ and $\tau_{geo}$ values have a modelled frequency dependence $\tau \propto \nu^{-4}$. At low frequencies the fits deviate most significantly from the $\tau_{geo}$ values.}
\label{fig:tauani}
\end{figure}

In the case of the anisotropic scatterer discussed here, the broadened profile is dependent on two characteristic scattering times, each with a frequency dependence $\tau_{x,y} \propto \nu^{-4}$, as described in Sec. \ref{sec:aniscat}. In general a more complex frequency dependence could be at play, since different observing frequencies will interact with anisotropic scattering ellipses of varying sizes.  Sampling different regions of the scattering screen in this way, can lead to additional frequency dependencies not included in our model.  

We show how an anisotropic scattering process will create apparent deviations from the expected $\tau$ spectrum when fitted with an isotropic scattering model. The scattering strengths are chosen to be $\sigma_{ax} = 3$ mas and $\sigma_{ay} =1$ mas at 1 GHz. That is to say the screen scatters 3 times more weakly in the chosen $y$-dimension than in the $x$-dimension.  Setting $\sigma_{ay} = 3$ mas will reproduce the isotropic scatterer considered in Sec. \ref{sec:isotau}. The scattering geometries are as before and the pulsar is the same 1 sec Gaussian profile from the previous section.

The simulated data are fitted with the train + DC method. The resultant profile fits for a selection of frequencies are shown in Fig.~\ref{fig:profani}. The corresponding $\tau$ spectrum is shown in Fig.~\ref{fig:tauani}. As before these are chosen to represent the mean profiles and spectrum after repeating the experiment with Gaussian noise of peak $\rm{SNR}=20$ multiple times. The mean and standard deviation of the spectral index as calculated after 1000 repetitions are given in Table~\ref{table:indexani}.

\begin{table}
\centering
\caption{The mean spectral indices as obtained by fits to profiles produced through an anisotropic scattering process.  The tabled spectral indices represent the average index as obtained from 1000 fitting procedures.}
\includegraphics[width = \columnwidth]{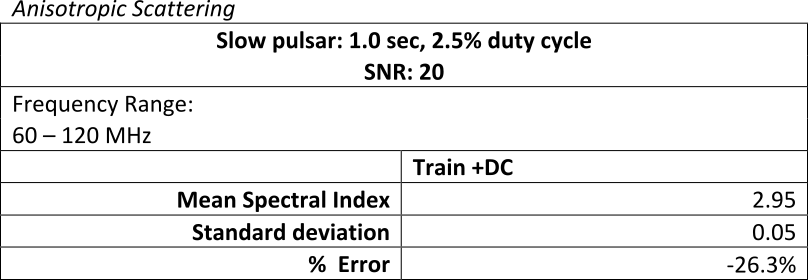}
\label{table:indexani}
\end{table}

In Fig.~\ref{fig:tauani} the input values of $\tau_1$ and $\tau_2$ (green solid and dashed lines) are shown as a reference.  Both have a frequency dependence with spectral index $\alpha = 4$. The geometric mean,

\begin{equation}
\tau_{geo} = \sqrt{\tau_x \tau_y}
\end{equation}

\noindent which would reduce the anisotropic scattering to isotropic scattering for $\tau_1 = \tau_2$, is computed from the input $\tau_1$ and $\tau_2$ and shown in yellow (stars). The single $\tau$ fit, shown in red, provides an estimate of the $\tau_{geo}$ value at mid frequencies. At low frequencies however the method fits deviate from $\tau_{geo}$ significantly.  Such an estimate of $\tau_{geo}$ will worsen as the degree of anisotropy increases.                                                                                                                                                                                                                                                                                                                                                                                                                                                                                                                                                                                                                                                                                                                                                                                                                                                                                                                                                                   

These results show that scatter broadened profiles caused by anisotropic scattering mechanisms require more sophisticated fitting models that can account for the degree of anisotropy. Such a model would make explicit use of eq.~\eqref{eq:ftani} in its description. For individual pulsars, screen anisotropy might be revealed through elongated images \citep{Brisken2010} or features of the secondary spectra, which will be instructive in determining the choice of fitting function, and in improving the fit.

\subsubsection{Flux-spectra}\label{sec:flux}

\begin{figure}
\includegraphics[width = 1.1\columnwidth]{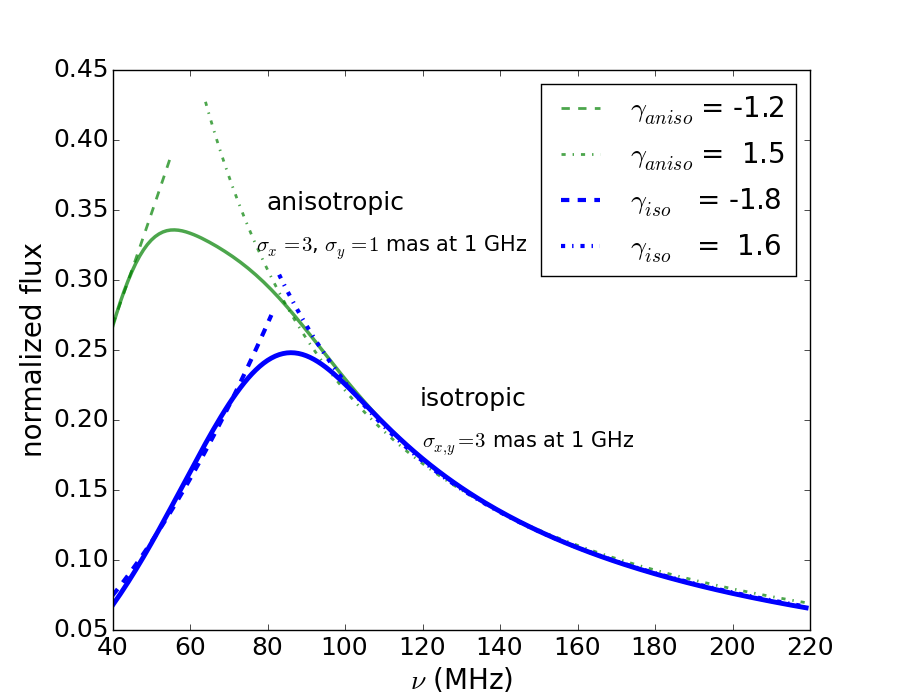}
\caption{Spectral turnovers in the pulsar flux spectra associated with infinite isotropic and anisotropic scattering screens. The turnovers at low frequencies are due to the raise in off-pulse baselines which leads to a loss in observable flux. These effects cause positive spectral slopes over the associated low frequency ranges. The spectrum related to anisotropic scattering here has a turnover at a lower frequency than the isotropic case. This is because the mean scattering strength in the chosen anisotropic case is weaker.}
\label{fig:fluxisoaniso}
\end{figure}

\begin{figure}
\includegraphics[width = \columnwidth]{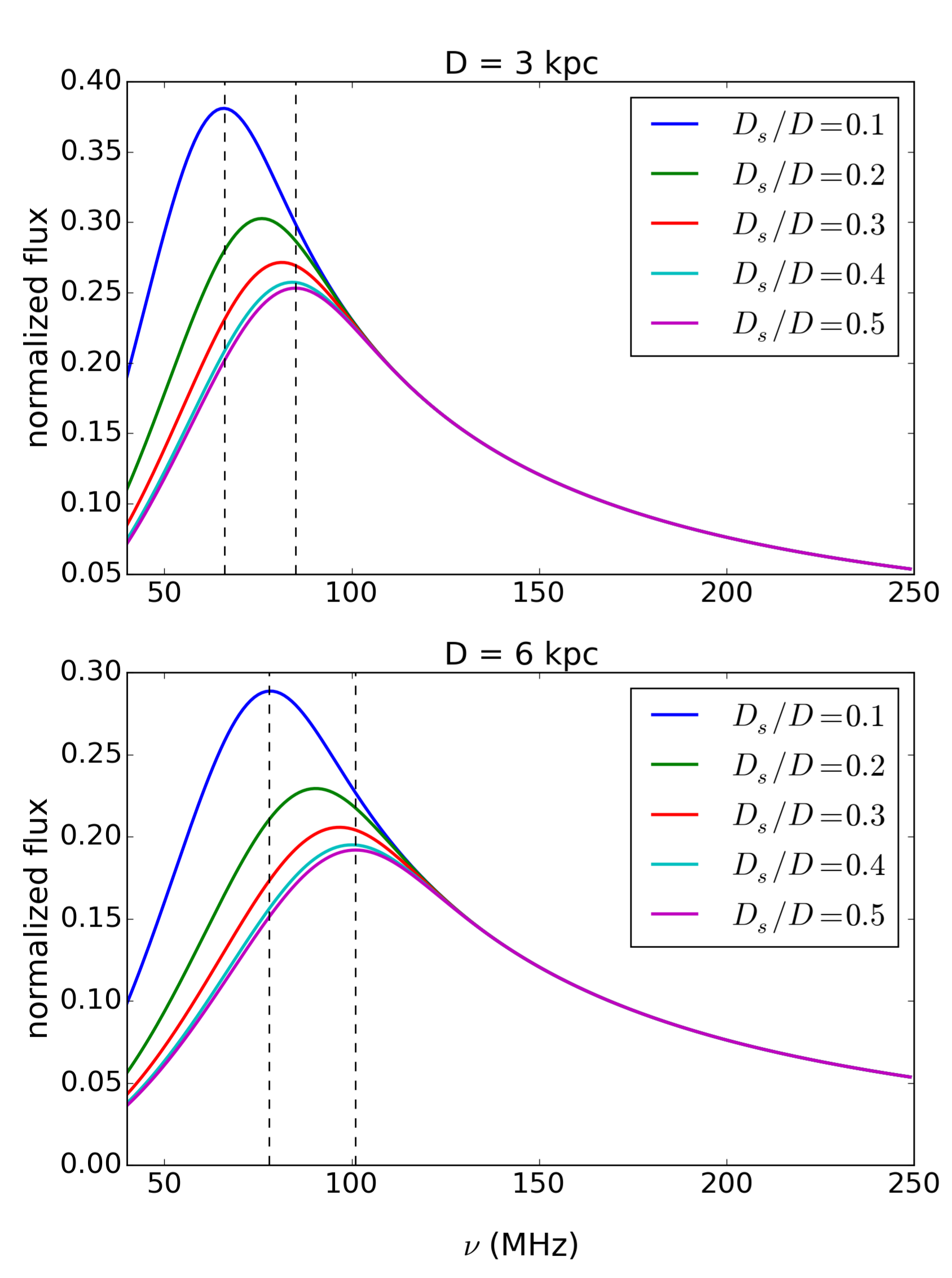}
\caption{Flux spectra plotted for different ratios of ${D_s}/D$ and overall distance $D$. The highest frequency at which a spectral turnover occurs increases from 86 MHz to 102 MHz as $D$ is increased from 3 to 6 kpc. Ever higher values of $D$ does not increase this value significantly ($D=12$ kpc has a turnover at $\sim$120~MHz).}\label{fig:isoDDs}
\end{figure}

The mean flux (${S_m}$) of a pulse profile is calculated by integrating the intensity of the profile averaged over the pulse period. This is illustrated schematically on the right-hand side of Fig.~\ref{fig:schmodel}, where the mean observable flux is shaded in blue. Scatter broadening that causes the power of pulses to be smeared into succeeding pulses will lower the observable flux, as intrinsic flux is lost to the raised baseline.  Since this is an artefact of the temporal domain, we can in principle circumvent the flux loss by imaging pulsars directly (e.g. \citealt{Dembska2015}).

We compute the flux for the setups in Sec. \ref{sec:tau} using the 1 sec pulsar, and show that a turnover in the flux spectrum occurs at low frequencies (high scattering). The location of this turnover is dependent on scattering mechanism and geometry. 

The mean flux values quoted here are calculated from the noiseless scatter broadened pulse shapes. It is normalised such that in the absence of the baseline being raised the lowest considered frequency (viz. 40 MHz)  will have a mean flux of unity. 

Fig.~\ref{fig:fluxisoaniso} show the turnovers in spectra for both the isotropic and anisotropic scattering setup. 
The turnover in the flux spectrum for the anisotropic scattering case occurs at a lower frequency and causes a smaller loss in flux since the mean scattering is weaker. The turnover also has an asymmetric shape.  

Fits to the high-frequency end of these spectra approximate the intrinsic spectral index of $\gamma = 1.6$. For low frequencies however the spectral turn-around leads to a positive slope such that the spectral index is estimated at $\gamma=-1.8$ for the isotropic case and $\gamma=-1.2$ for the anisotropic case.

In Fig.~\ref{fig:isoDDs} we illustrate how changing the scattering geometry impacts on the flux spectra. We show that as the position of the screen (${D_s}$) is moved for a fixed overall distance ($D$) the frequency at which we expect to observe spectral turnover due to baseline effects will change. This phenomenon is symmetric about the mid-point (${D_s}/{D}= 0.5$) such that for example the spectra for ${D_s}/{D}= 0.8$ and 0.2 are equal. The observed scattering effects are at its maximum for ${D_s}/{D}= 0.5$ and therefore the frequency at which flux loss is first observed will be highest for a midway screen. 

The overall distance ${D}$ is increased from 3~kpc to 6~kpc from the top panel to the bottom panel of Fig.~\ref{fig:isoDDs}. This overall increase in distance increases the scattering and subsequently the frequency at which spectral turnover appears. Simultaneously as $D$ is increased the frequency window within which spectral turnover takes place for all plotted ratios of ${D_s}/D$ (indicated with dashed lines) grows slowly.  This result shows that for distances within our Galaxy and a scattering strength of $\sigma_\theta = 3$~mas at 1~GHz, as taken from \citet{CordesLazio2001}, turnovers in mean flux spectra will occur below $100$~MHz.  Spectral turnovers at higher frequencies are either due to different scattering mechanisms (deviations from isotropically Gaussian scattering) or different scattering strengths than the typical value used here. 
However since a wide range of scattering strengths are expected, it is unclear whether or not a coherent picture is expected throughout the population of pulsars with measured spectra.

At very large distances, $D > 200$~kpc,  (while still using the above scattering strength) the simulated pulse is completely scattered out and a turnover in the flux spectrum no longer detectable in this way.

\subsection{Effects of screen geometry}\label{sec:geoeffects}

In Sec. \ref{sec:numbroad} we discussed computing the temporal broadening functions ($f_t$s) of truncated screens by means of a ray-tracing code. We expect these finite screens to cause a loss in the observable flux of the pulsar. Furthermore we investigate how the shape of an average pulse profile, and consequently the $\tau$ spectrum, is impacted by the presence of such a screen.

By defining the screen in terms of its height and width we can create both square screens that will have a comparable impact to a circular screen or rectangular (elongated) screens. In the theoretical limit where screens become infinitely more long than wide (i.e. 1D filaments) the broadening function will tend to,

\begin{align}
f_{t} &= e^{-t/\tau}/(\sqrt{\pi t \tau}) U(t).\label{eq:ft1D}
\end{align}

\noindent We stick to screens where the ratio of its dimensions has not yet reached this theoretical limit.

\subsubsection{Profile effects}

\begin{figure}
\centering
\includegraphics[width = \columnwidth]{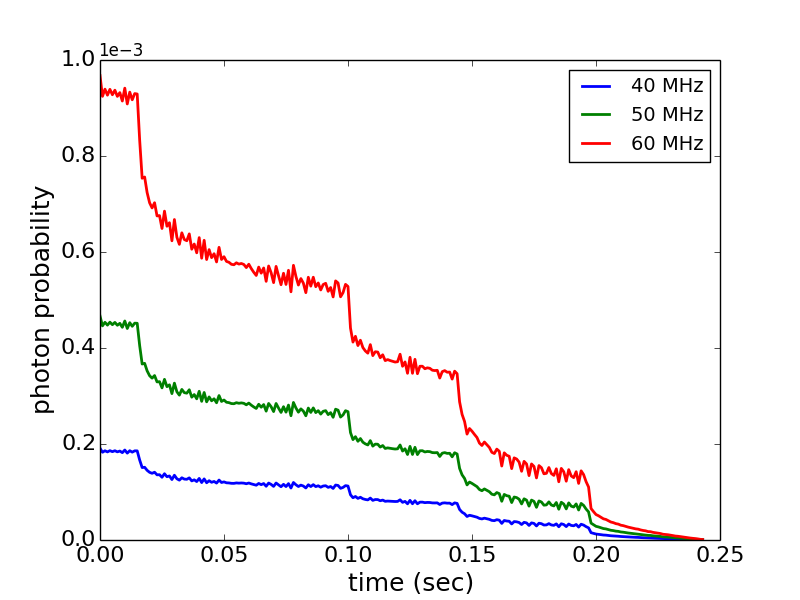}
\caption{The broadening function associated with a truncated scattering screen of size 400 AU by 600 AU, with the pulsar offset to the centre of the screen as discussed in the text. The functions are generated numerically by means of a ray tracing experiment that calculates the probability of photons arriving at the observer at a given time.} 
\label{fig:broadtrunc}
\end{figure}

The broadening function for a source at a distance of \mbox{${D}= 3$ kpc} with a finite isotropic scattering screen halfway between the source and the observer is presented in Fig.~\ref{fig:broadtrunc}. The modelled screen has size 400 AU by 600 AU, with the pulsar offset to the middle of the screen by 100 AU horizontally and 50 AU vertically. The scattering strength is as before, $\sigma_a = 3$ mas at 1GHz. 

The dips in the broadening function are associated with the four edges of the screen and show the associated loss in flux at different frequencies. In this instance we see dips at $0.02, 0.10, 0.15$ and $0.20$ s associated with distances to the edge of $100, 250, 300$ and $350$ AU. 

\begin{figure}
\centering
\includegraphics[width = \columnwidth]{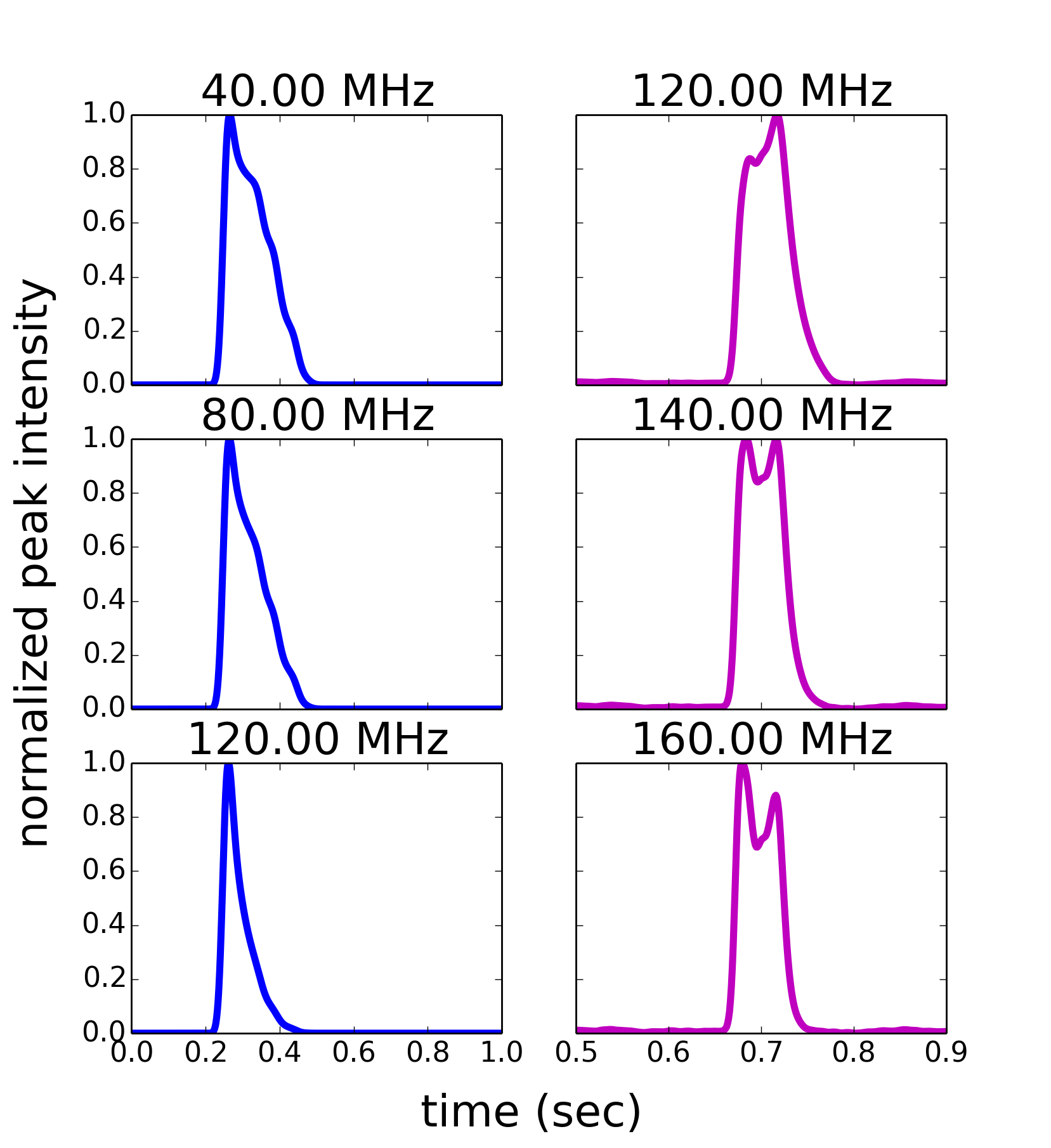}
\caption{\textit{Left:} Convolutions of an intrinsic Gaussian pulsar profile with the temporal broadening functions shown in Fig.~\ref{fig:broadtrunc}. The scattering screen has size 400 AU by 600 AU, offset with respect to the line of sight as described in the text. The distances are \mbox{$D=3$~kpc} and $D_s=D/2$~kpc. At 40~MHz and 80~MHz  the effect of a truncated screen is seen as bumps and an overall steepening of the profile shape. At 120~MHz the profile starts to resemble the typical exponential scattering tail. \textit{Right:} Broadened pulses using B1237+25 as a profile template and a screen of size 200 AU by 200 AU, centred on the line of sight, with ${D}=0.85$~kpc and \mbox{$D_s=D/2$}. The relative heights between the two peaks change as the observing frequency changes. All profiles are plotted normalised to their height.}
\label{fig:truncprofnonoise}
\end{figure}

The numerical broadening functions are convolved with the same ${P}= 1.0$ s  pulsar as used for the infinite scattering screens in Sec. \ref{sec:resinf}. A selection of the resulting broadened profiles are shown on the left-hand side of Fig.~\ref{fig:truncprofnonoise}. 

The broadening effects can be split into three regimes. At high frequencies (above 210 MHz, not shown) the effect of scattering is negligible and the observed pulse resembles the intrinsic Gaussian pulse. At mid-frequencies (here 120 -- 210 MHz) the pulse shapes have the typical exponential tail as would be associated with an infinite scattering screen. At low frequencies (below 120 MHz) the finite nature of the scatterer becomes visible in the form of bumps associated with the loss in flux at various path lengths (or times).  These pulse shapes have a much steeper trailing edge than normally associated with a scatter broadened pulse shape. 

As an example of these effects on a more complicated profile we use a high frequency profile of B1237+25 as a multi-component template and consider how this template would change with frequency when it is scattered by a truncated screen. In our simulations we use the pulsar's period ($P=1.38$~s) and distance (${D}= 0.85$ kpc), while acknowledging that the pulsar itself, being a high galactic latitude source, is not an ideal candidate for scattering studies. We consider the impact of an isotropic screen, size 200 by 200~AU, placed midway along the line of sight.  The scattering strength is as before. The resulting broadened pulse shapes at several frequencies are shown on the right-hand side of  Fig.\ref{fig:truncprofnonoise}.

Note how the truncation effect of the screen can cause
the height of the two profile peaks to change relative to each other. Between 110 and 130~MHz the left peak is lower than the right one, at 140 MHz the peaks are roughly equal, and beyond 140~MHz the left peak is the higher peak. The change in features in these profiles could easily be misinterpreted as features of the intrinsic pulse shape.

The above scattering setups are theoretical constructs, however by investigating the impact of truncated screens on both the $\tau$ and flux spectra, we can build a more complete picture of all the possible sources of deviations to the otherwise theoretically predicted spectral trends.

\subsubsection{Effects on $\tau$ spectra}

\begin{figure*}
\centering
\includegraphics[width = \textwidth]{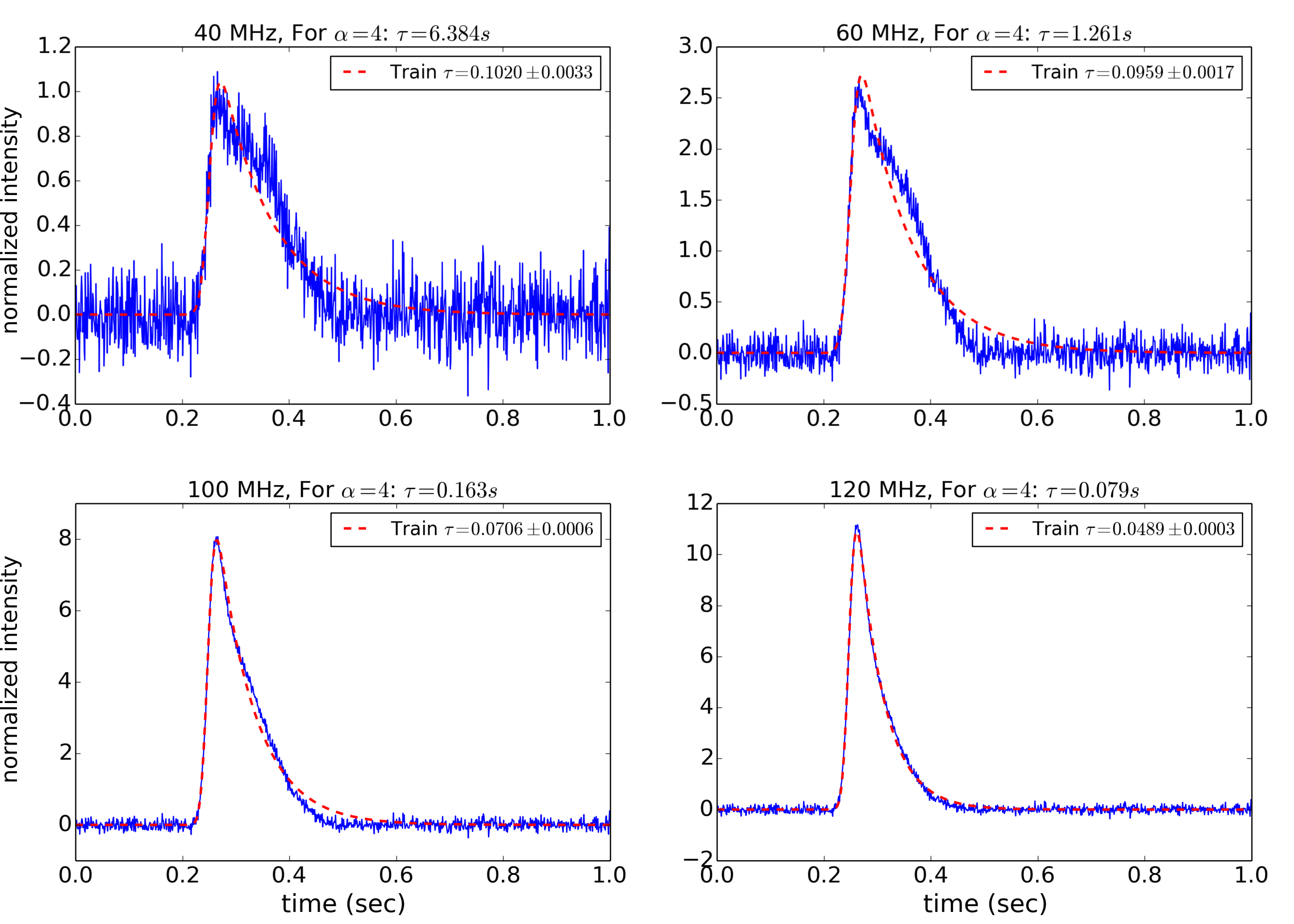}
\caption{Profiles as generated by the truncated broadening function of Fig.~\ref{fig:broadtrunc} with added Gaussian noise. The profiles plotted here are normalised relative to the peak of the profile at 40 MHz (which has suffered the greatest flux loss).  The input $\tau$ values at different frequencies are given, along with the values obtained from the fits with the train method. In all instances the $\tau$ values are underestimated by more than 45\%. }
\label{fig:truncprofiles}
\end{figure*}

\begin{figure}
\includegraphics[width = \columnwidth]{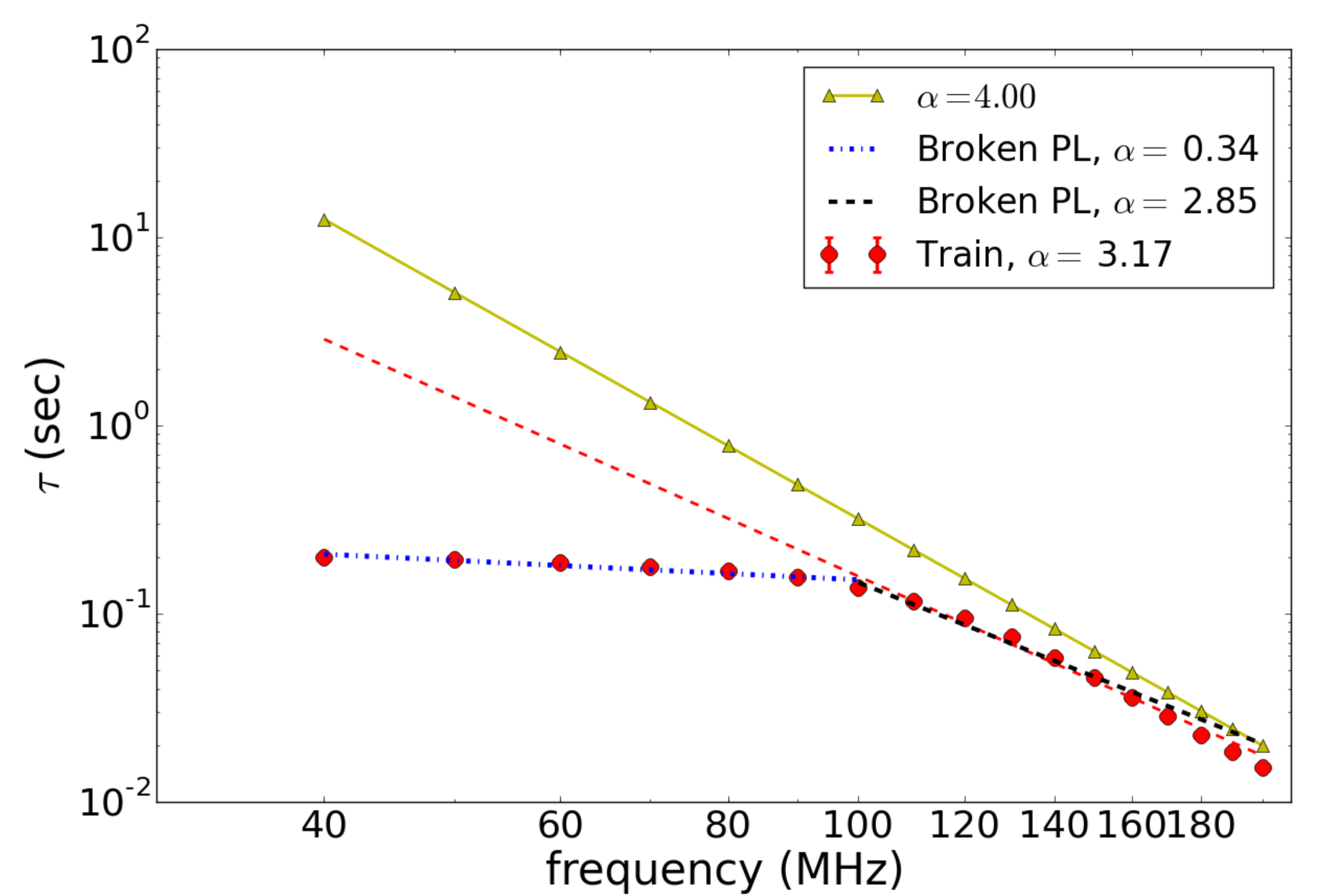}
\caption{The $\tau$ spectrum associated with a truncated screen. At low frequencies (high scattering) the $\tau$  values as obtained by the train method (red circles, with error bars smaller than the marker) greatly underestimates the input scattering (yellow triangles). This is due to the fact that at large wavelengths the scattering process will become sensitive to the edges of the screen leading to imprints in the temporally broadened pulse profiles for which our method does not account. The short black (dashed) and blue (dot-dash) lines show a broken power law fit with spectral indices as indicated in the legend, and the long red (dashed) line is the weighted power law fit of all the obtained $\tau$ values.}
\label{fig:tautrunc}
\end{figure}

To estimate the $\tau$ spectra we added Gaussian noise to the simulated profiles, after which we fitted them with the train method of Sec. \ref{sec:fittech}. (Since the truncation prevents wrap around profiles here we do not have to include a DC offset parameter.) The SNR value is modelled to scale with the mean flux of the profile, and is therefore notably smaller at low frequencies for which the flux loss is greater. (We discuss the flux spectrum in more detail in the next section.) The resulting profile fits for a selection of frequencies are shown in Fig.~\ref{fig:truncprofiles}. 

The corresponding $\tau$ spectrum is shown in Fig.~\ref{fig:tautrunc}.  From the $\tau$ spectrum it is clear that a model based on the assumption of an infinite screen, greatly underestimates the true value of $\tau$. A weighted fit to all the obtained $\tau$-values produces a spectral index of \mbox{$\alpha=3.17$}. 
If we identify a single frequency at which the observations appear to become sensitive to the edges of the screen, we can fit the spectrum with two power laws instead. Using 100 MHz as this breakpoint, we find that \mbox{$\alpha =2.85$} above this frequency. Below 100 MHz we see a flattening of the $\tau$ spectrum, such that the obtained spectral index here is  \mbox{$\alpha=0.34$}.

\subsubsection{Effects on flux-spectra}

\begin{figure}
\includegraphics[width=1.1\columnwidth]{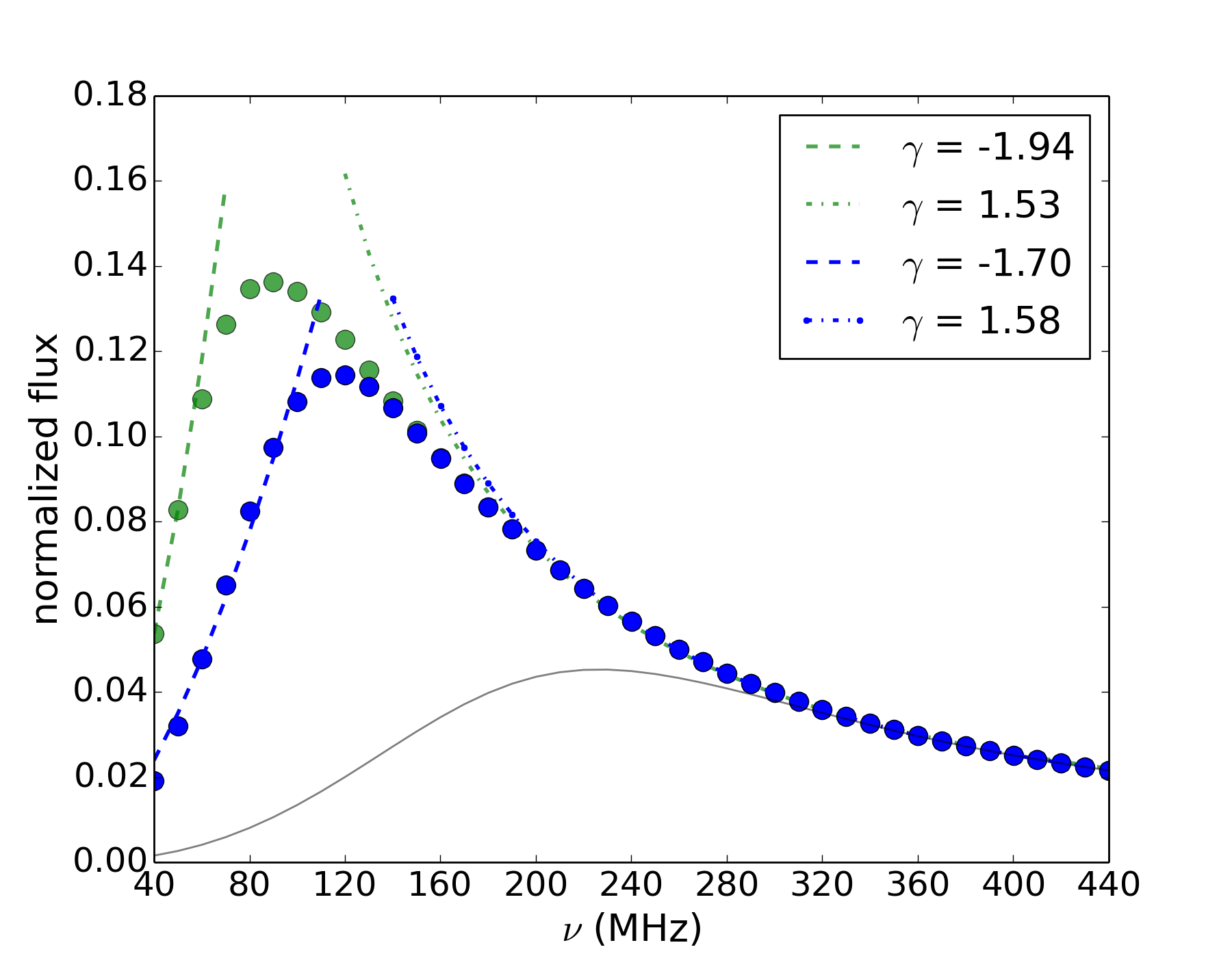}
\caption{The flux spectrum of the truncated scattering screen setup discussed in the text is shown in blue. The turnover in the spectrum is due to the longer wavelengths being emitted beyond the boundaries of the screen such that not all of the pulsar's flux fill be received back at the detector. The equivalent anisotropic case for a screen that scatters 3 time more weakly in one dimension is presented in green. The turnover in this case will occur at a lower frequency compared to the stronger isotropic scatterer. The grey line shows the flux spectrum for a smaller screen (100 by 200~AU). Decreasing the screen size leads to turnovers at higher frequencies.}
\label{fig:truncflux}
\end{figure}

We expect to measure a loss in the observable flux for a pulsar behind a truncated scattering screen of such size that at low frequencies not all the photons from the pulsar will be refracted back to our line of sight. To model this loss we calculate the probability that a photon hitting the truncated screen at some angle and location on the screen, will reach the observer. We then compare this to the probability associated with a sufficiently infinite screen. \textit{Sufficiently infinite} is defined in terms of the width of the Gaussian distribution from which the scattering angle is drawn. A distribution that runs from -$5~\sigma_{\theta}$ to $5~\sigma_{\theta}$ is considered to be an approximation of an infinite screen. 

The ray-tracing code is normalised such that the temporal broadening function of a sufficiently infinite screen at the lowest studied frequency produces an impulse flux of unity. This means when convolving a simulated pulse with an infinite screen the resulting flux spectrum will equal the intrinsic flux spectrum of the pulsar with ${S_m} = 1$ at the lowest observed frequency (provided that no flux is lost due to high levels of scattering that have lead to wrap around scattering tails).

The flux spectrum in Fig.~\ref{fig:truncflux}, associated with isotropic scattering, shows a turnover close to $120$ MHz. Above this frequency the screen is effectively infinite such that the flux of the pulsar behind it is conserved, and the intrinsic pulsar spectrum with $\gamma = 1.6$ becomes visible. Below $120$ MHz, the spectral index is positive as more flux is lost towards lower frequencies (longer wavelengths) when the scattering process becomes sensitive to the edges of the scattering screen.  

Fig.~\ref{fig:truncflux} also contains the flux spectrum for a truncated  anisotropic scatterer of the same size and location. The anisotropy is created by again choosing that the screen scatters 3 times more weakly in one dimension. Since this leads to an altogether weaker scatterer the loss in flux is less severe and we observe the onset of flux loss at a lower frequency that for the isotropic example.

The turnovers in both Figs. \ref{fig:fluxisoaniso} and \ref{fig:truncflux} occur at frequencies below 150 MHz for the current scattering setup. Pushing the overall distance (D) to 6 kpc in Fig.~\ref{fig:isoDDs} did not greatly increase the frequency at which the spectral turnover due to baseline effects occur. An alternative way to produce turnovers at higher frequencies, while keeping the scattering strength, could be by means of ever smaller scattering clouds. Fig.~\ref{fig:truncflux} includes an example (as a grey line) of a smaller screen with a higher turnover frequency. 

\section{Discussion}\label{sec:discussion}
Reflecting on the results in the Sec. \ref{sec:results} we make the following statements.

In Sec. \ref{sec:end2end} we show that errors related to estimating the 
off-pulse baseline can be circumvented by treating the baseline level (DC offset) as a parameter for which our methods fit. This ensures that the estimated $\tau$ value is centred on the true simulated value for $\tau$. 

Table \ref{table:one} summarises the impact of using different methods to estimate the spectral indices. Errors in $\tau$ increase towards lower frequencies (or higher scattering) for all the fitting methods. Given a pulse shape and a duty cycle, $\tau$ in millisecond pulsars will become comparable to the pulse period at higher frequencies than for normal pulsars. 

Larger errors in $\tau$ lead to larger errors in $\alpha$. Our simulations show that at high levels of scattering there is an underlying asymmetry in the measurement uncertainties of $\tau$. Correctly accounting for the asymmetry results in a more accurate measurement of $\alpha$. The impact of this correction is seen for the millisecond pulsar case, where the mean estimate of $\alpha$ changes from 3.87 to 4.01 over the low frequency range, while the standard deviation in the estimate simultaneously drops from 0.32 to 0.17.  Having introduced this correction, the $\alpha$ values for all our simulated data have a mean value within 0.3\% of the true value, and a standard deviation below 0.2.
 
Increasing the SNR also leads to a decrease in the standard deviation of $\alpha$ estimates. In the case of the 1.0 sec pulsar at 60 -- 120 MHz, an increase of SNR from 10 to 20 to 50 decreases the standard deviation by a factor of 2 and 2.25, as shown in Table \ref{table:snr1050}. 

The EMG method in this paper portrays the most basic model, where the wrap around of a scattering tail is not modelled. For such an implementation deviations from the theoretically expected values at low frequencies can be as large as 30\% - 70\% for the pulsars and scattering scenarios we simulate. The failure of this model is made explicit to encourage the revision of fits in the literature where wrap around scattering tails may not have been modelled, e.g. Fig.~1 in \citet{Lewandowski2013}.

We show in Fig.~\ref{fig:fcfm} that it is important to attribute the measured $\tau$ values to the associated monochromatic frequency (${f_m}$, computed from the central frequency and bandwidth of the observation as in eq. \ref{eq:fmfc}) to obtain accurate estimates of the spectral index ($\alpha$) from a power law fit.

We investigate the scatter broadening of pulsars due to anisotropic scattering, which we model as an asymmetric Gaussian distribution, and show that fitting these profiles with isotropic models leads to systematic errors in the measurement of $\tau$. Assuming the scattering strength in one dimension to be weaker than what we have considered as typical in our simulations, such fits will lead to apparent $\tau$ spectra with $\alpha< 4$ (in our example $\alpha = 2.95$). Improvements to such fits will rely  on using more appropriate broadening functions, such as eq. \eqref {eq:ftani} or \eqref{eq:ft1D}, in the fitting model. We are currently investigating the use of anisotropic fitting techniques to analyse scattered profiles from LOFAR observations. Given the recent evidence for highly anisotropic scattering our simulations show that the measured $\alpha$ (and goodness of fit for $\tau$) could serve as a diagnostic in the time domain for the type of scattering screen. 

From the simulated truncated screens in Sec. \ref{sec:geoeffects}, we show that an isotropic thin screen with dimensions 400~AU by 600~AU midway between the Earth and a pulsar at 3~kpc, changes the shape of the pulse profile with frequency and leads to a flattened $\tau$ spectrum below 200~MHz. In Fig. \ref{fig:tautrunc} the spectral index was estimated to be $\alpha = 0.34$ below 100~MHz. Adjustments to the modelled screen size and location will impact the shape and features of the $\tau$ spectra, resulting in possible breaks. 

For a complex pulse profile the frequency evolution due to this kind of scattering may change the component amplitude ratio. We use the profile of B1237+25 to demonstrate this effect, although this pulsar is not observed to exhibit such behaviour. The effects caused by scattering can in principle be disentangled from the intrinsic emission, as it is unlikely that both will have the same dependence on frequency. 

\begin{figure}
\includegraphics[width = \columnwidth]{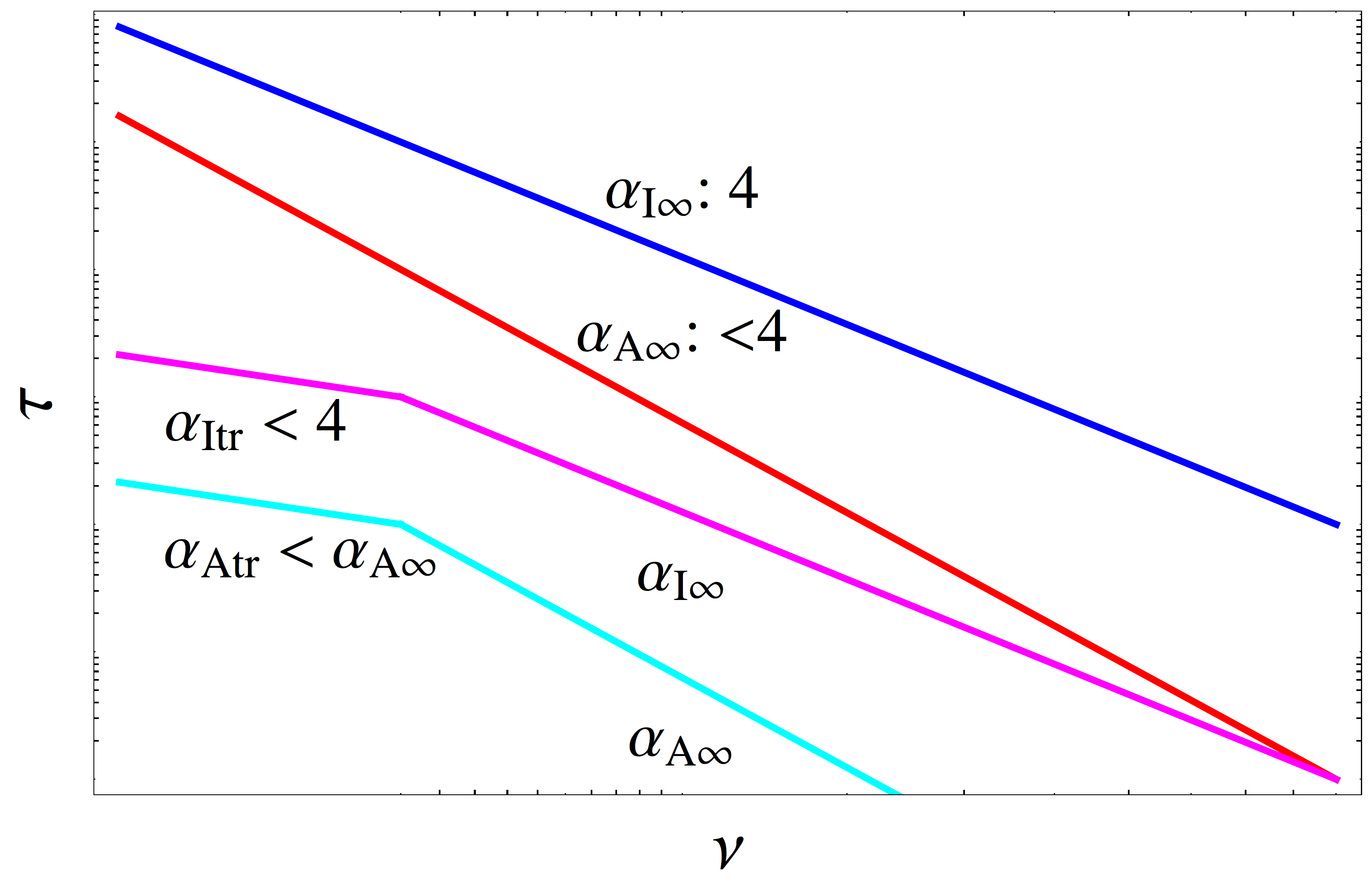}
\caption{A cartoon summary of the deviations from the theoretical $\tau$ spectrum and their underlying causes. Subscripts $I$ and $A$ refer to isotropic and anisotropic scattering mechanisms, and $\infty$ or $tr$ to either infinite or truncated thin screens. The anisotropic scatterer shown here scatters more weakly in one dimension than what is considered a typical isotropic scattering strength.}
\label{fig:scatcart}
\end{figure}

Fig.~\ref{fig:scatcart} provides a cartoon style summary of deviations from the theoretical $\tau$ spectrum we use (a power law spectrum with spectral index 4) and the underlying causes.  Subscripts $I$ and $A$ refer to isotropic and anisotropic scattering mechanisms, and $\infty$ or $tr$ to either infinite or truncated thin screens.

For the range of scattering mechanisms and geometries discussed above, we also investigated the impact on pulsar flux spectra. Observationally, pulsar flux spectra have steep negative spectral indices, typically within the range $\gamma = -1.8$ to $-1.4$ \citep{Lorimer1995,Maron2000,Bates2013}.  Spectral turnovers at low frequencies are a known feature of the population of slow pulsars (e.g. \citealt{Lorimer1995}).  Thermal absorption by dense surrounding environments (e.g. \citealt{Rajwade2016}) may be partially responsible for spectral turnovers, which may also be intrinsic to the radio emission of the pulsar. Some evidence suggests that the efficiency of the radio emission shuts off at lower frequencies \citep{Malofeev2000}. Alternatively, under the assumption that the lower frequencies are emitted higher above the neutron star surface than the higher frequencies \citep{Radhakrishnan1969}, the intersection of the line of sight with the radio emitting regions in the magnetosphere will also be frequency dependent, with the possibility of the line of sight gradually missing the emitting regions altogether at low frequencies. A small group of pulsars show turnovers at much higher frequencies. These are known as the gigahertz peaked spectrum (GPS) pulsars. The GPS turnovers are also attributed to either the radio emission process becoming inefficient above a threshold frequency, or dense ISM environments \citep{Kijaketal2007,Kijak2011}. 

Here we concentrate on the effects of scattering on the observed flux, which should occur in addition to the above mechanisms. In our simulations we show that high levels of scattering by infinite screens, result in a measurable integrated flux loss, due to the rise in the off-pulse baseline. In the case of slow pulsars this will lead to spectral turnovers typically below 100 MHz, as shown in Fig. \ref{fig:isoDDs}. The turnover frequency due to this effect in millisecond pulsar flux spectra, would be higher than for slow pulsars. Older studies of millisecond pulsar flux spectra have mostly not found turnovers \citep{Kuzmin2000, Kuzmin2002}, however more recent work by \citet{Kuniyoshi2015} have observed turnovers in MSP flux spectra, albeit in imaging observations.This could point to anomalous scattering by e.g. finite scattering clouds. The lack of observed spectral turnovers from time-domain studies suggest that most MSP profiles are not substantially scattered in current low frequency observations. 

Fig. \ref{fig:fluxisoaniso} shows the difference in turnover frequencies for an infinite isotropic and anisotropic scattering screen. In our example of an anisotropic screen, the spectral turnover occurs at $\sim50$~MHz. The turnover frequency for the corresponding isotropic case was $\sim90$~MHz, i.e. higher by a factor of $\sqrt{\sigma_{a_x} \sigma_{a_y}{\color{white}^i}}/\sqrt{\sigma_i^2}$ (where subscripts $a$ and $i$ refer to anisotropic and isotropic respectively and $(x,y)$ are the two dimensions describing the thin screen, with $\sigma_i = \sigma_{a_x} = 3$ and $\sigma_{a_y}$ = 1 mas at 1GHz).

Similar spectral turnovers can be caused by the loss of flux associated with truncated scattering screens. In Fig.~\ref{fig:truncflux} we show that a finite screen of size 400~AU by 600~AU at the same location as the infinite screen, would lead to turnover frequencies in approximately the same range, with the isotropic flux spectrum peaking at 125~MHz and the anisotropic spectrum at 90~MHz. Keeping the scattering strength constant, the turnover frequency can be pushed up to 230~MHz for a screen of size 100~AU by 200~AU, centred on the line of sight. In this paper, we have concentrated on screens with infinitely small inner scales, i.e. the scales of structure in the screen that result in the largest scattering angles. A larger inner scale, as discussed in \citet{RickettJohnston2009}, would result in a maximum scattering angle, similar to the case of a screen of finite size, and hence similar spectra for $\tau$. The crucial difference between the two is that the existence of an inner scale does not reduce the total observed flux, as is the case for a finite screen.

To reproduce GPS spectra with our modelling would require a fine tuning of the truncated screen size and or the simulated scattering strength. Studying the correlated changes in flux spectra, pulsar profiles and their associated $\tau$-spectra with frequency can reveal the complex interplay between the phenomena that set the observed pulse profile and its dependence on frequency, and should lead to a better understanding of the effects of each. 

\section{Conclusions}
We have showed that our train+DC method is an efficient forward fitting technique for scattered profiles. Our investigations show flatter $\tau$ spectra (than what is theoretically expected) when using an isotropic model to fit profiles shaped by an anisotropic scattering mechanism.  Flatter $\tau$ spectra can also be the result of low frequency profile shape changes caused by a truncated scattering screen. Over a large enough sampled frequency range, such a $\tau$ spectrum will exhibit a break. Analysing the impact of scattering on flux spectra, has revealed turnovers at low frequencies due to baseline effects. Similarly truncated screens can cause turnovers in flux spectra. These turnovers are a function of the screen size and its scattering strength. 

We anticipate new data from instruments such as LOFAR and the MWA in the frequency range studied here. This paper will serve as a framework to compare real data to the simulated results obtained here. Our analysis of scattering imprints is by no means complete, as we have only investigated simple profile shapes and Gaussian scattering mechanisms. However, our results indicate where attention should be paid in terms of creating accurate fitting methods as well as describing possible sources of deviations to theoretically predicted results. 

\section{Acknowledgements}
M. Geyer is a Commonwealth Scholar funded by the UK government.  The authors also express their gratitude to SKA South Africa for their continued support.
We would like to thank Jayanth Chennamangalam and Mark Walker for valuable discussions and insights relating to this work. We thank the anonymous reviewer for their valuable comments to improve the quality of the paper.

\bibliographystyle{mn2e.bst}
\bibliography{ScatteringBibDesk}

\end{document}